\pdfoutput = 1
\documentclass[aps,prb,twocolumn,showpacs,nobalancelastpage,superscriptaddress,10pt]{revtex4-1}
\usepackage{tabulary}
\usepackage{graphicx}
\usepackage{color}
\usepackage{amsmath}
\usepackage{amsfonts}
\usepackage{bm}
\usepackage{enumerate}
\usepackage{array}

\newcommand{\Ket}[1]{\vert  #1 \rangle}
\newcommand{\Bra}[1]{\langle #1 \vert}
\newcommand{\MatEl}[3]{\langle #1 \vert #2 \vert #3 \rangle}

\newcommand{\be}{\begin{equation}}
\newcommand{\ee}{\end{equation}}
\newcommand{\bea}{\begin{eqnarray}}
\newcommand{\eea}{\end{eqnarray}}

\newcommand{\ket}[1]{\ensuremath{|#1\rangle}}
\newcommand{\braket}[2]{\langle#1|#2\rangle}

\newcommand{\avg}[1]{\ensuremath{\langle#1\rangle}}

\newcommand{\mc}[1]{\mathcal{#1}}
\newcommand{\ex}[1]{\langle#1\rangle}
\newcommand{\qb}[2]{\langle#1|#2\rangle}

\renewcommand{\vec}[1]{{\bm #1}}
\renewcommand{\epsilon}{\varepsilon}

\global\long\def\bra{\langle}

\global\long\def\ket{\rangle}
\global\long\def\kett#1{|#1\rangle}

\global\long\def\s#1{\mathcal{#1}}
\global\long\def\ex#1{\bra#1\ket}
\global\long\def\p#1{\left(#1\right)}
\global\long\def\qb#1#2{\langle#1|#2\rangle}

\global\long\def\dag{\dagger}

\renewcommand{\vec}[1]{{\bm #1}}

\global\long\def\s#1{\mathcal{#1}}
\global\long\def\bra{\langle}

\global\long\def\ket{\rangle}
\global\long\def\kett#1{|#1\rangle}

\global\long\def\s#1{\mathcal{#1}}
\global\long\def\ex#1{\bra#1\ket}
\global\long\def\p#1{\left(#1\right)}
\global\long\def\qb#1#2{\langle#1|#2\rangle}

\global\long\def\dag{\dagger}

\global\long\def\a{\alpha}

\begin{document}

\title{Controlled Population of Floquet-Bloch States via Coupling to Bose and Fermi Baths}

\author{Karthik I. Seetharam}
\affiliation{Institute for Quantum Information and Matter, Caltech, Pasadena,
California 91125, USA}
\author{Charles-Edouard Bardyn}
\affiliation{Institute for Quantum Information and Matter, Caltech, Pasadena,
California 91125, USA}
\author{Netanel H. Lindner}
\affiliation{Physics Department, Technion, 320003 Haifa, Israel}
\affiliation{Institute for Quantum Information and Matter, Caltech, Pasadena,
California 91125, USA}
\author{Mark S. Rudner}
\affiliation{Niels Bohr International Academy and Center for Quantum Devices, University of Copenhagen, 2100 Copenhagen, Denmark}
\author{Gil Refael}
\affiliation{Institute for Quantum Information and Matter, Caltech, Pasadena,
California 91125, USA}

\begin{abstract}

External driving is emerging as a promising tool for exploring new phases in quantum systems.
The intrinsically non-equilibrium states that result, however, are challenging to describe and control.
We study the steady states of a periodically driven one-dimensional electronic system, including the effects of radiative recombination, electron-phonon interactions, and the coupling to an external fermionic reservoir. Using a kinetic equation for the populations of the Floquet eigenstates, we show that the steady state distribution can be controlled using the momentum and energy relaxation pathways provided by the coupling to phonon and Fermi reservoirs. In order to utilize the latter, we propose to couple the system and reservoir via an energy filter which suppresses photon-assisted tunneling. Importantly, coupling to these reservoirs yields a steady state resembling a band insulator in the Floquet basis. The system exhibits incompressible behavior, while hosting a small density of excitations. We discuss transport signatures, and describe the regimes where insulating behavior is obtained. Our results give promise for realizing Floquet topological insulators.

\end{abstract}


\maketitle



The availability of coherent driving fields such as lasers opens many exciting possibilities for controlling quantum systems.
In particular, the recent realization that the topological characteristics of Bloch bands can be modified through periodic driving~\cite{Niu2007,Oka2009,Lindner2011, KBRD} sparked a wave of proposals~\cite{Jiang2011,Kitagawa2011,Gu11, Lindner2013,Rudner2013,Delplace2013,Kundu2013,Kundu2014,Grushin2014,Goldman2014} and experiments~\cite{Kitagawa2012,Rechtsman2013,Wang2013,Jotzu2014,JimenezGarcia2014} to realize various types of
``Floquet topological insulators'' in solid state, atomic, and photonic systems.
Here topology emerges in the basis of Floquet states, time-periodic eigenstates of the driven system's single-particle
evolution operator~\cite{Shirley1965,Sambe1973,Dittrich1998}.

\begin{figure}[t]
    \begin{center}
        \includegraphics[width=0.85\columnwidth]{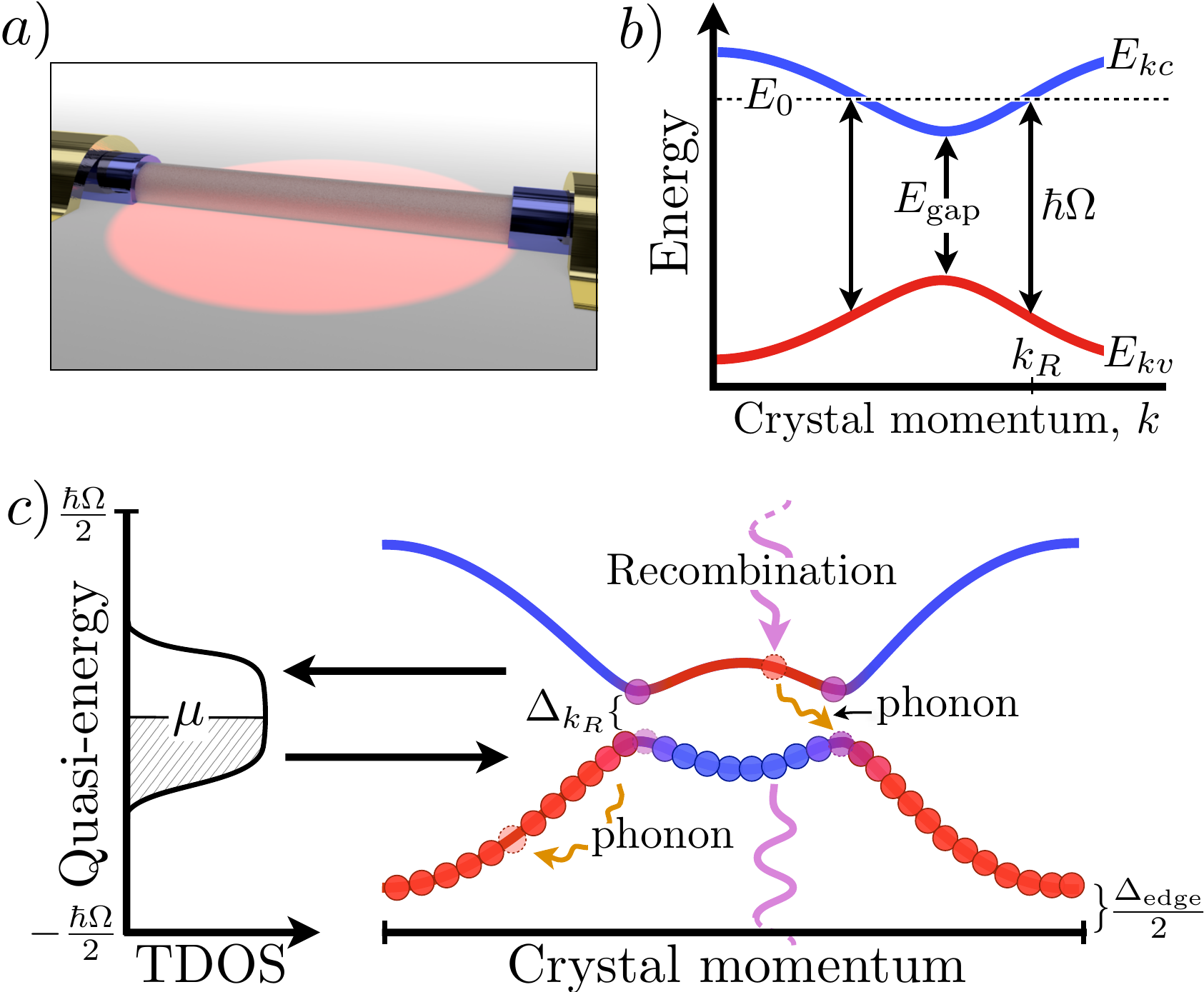}
        \caption{Carrier kinetics in a Floquet-Bloch system coupled to Bose and Fermi reservoirs.
a) One dimensional semiconductor wire coupled to an energy-filtered fermionic reservoir.
Energy filtering is achieved by coupling the system and reservoir via a deep impurity band in a large bandgap semiconductor.
b) Band structure of the non-driven system.
The driving field photon energy $\hbar\Omega$ exceeds the bandgap $E_{\rm gap}$,
causing resonant coupling at crystal momentum values $\pm k_R$.
c) Floquet band structure, indicating the character of the Floquet band in terms of the original conduction (blue) and valence (red) bands.
Coupling to acoustic phonons mediates electronic momentum and energy relaxation (orange arrows), while radiative recombination scatters electrons vertically between conduction and valence band like states (purple arrow). At half filling, the steady state resembles that of an insulator with a small density of excited electrons and holes.
}
\vspace{-0.4 in}
        \label{fig:setup}
    \end{center}
\end{figure}

Floquet states provide a convenient basis for describing the evolution of driven systems, in many ways analogous
to the Hamiltonian eigenstates of non-driven systems.
However, the powerful thermodynamic rules that govern the level occupations of static systems in thermal equilibrium in general {\it cannot} be directly translated into the inherently non-equilibrium context where Floquet states are defined \cite{Kohn2001,Hone2009, Wustmann2010,Kohler2005}. Photon-assisted scattering processes in which energy is exchanged with the driving field produce heating and violate the reversibility conditions that give rise to the Boltzmann distribution in equilibrium~\cite{Bilitewski2014}.
This crucial difference brings up many intriguing and important questions about the steady-state physical properties of open Floquet-Bloch systems.
In particular, in order to realize the promise of non-equilibrium topological phenomena, one of the major outstanding problems 
is to identify which types of systems, baths, and system-bath couplings can lead to non-equilibrium steady states enabling Floquet topological insulators to exhibit behaviors similar to those of their equilibrium counterparts~\cite{Dehghani2014, Dehghani2014b}.

Floquet-Bloch steady-state distributions are currently known for certain special cases.
A closed driven system inevitably heats up to a maximal entropy (infinite temperature) steady state~\cite{Lazarides2014, DAlessio2014}, in the absence of many-body localization~\cite{Ponte2014, Abanin2014}.
In contrast, an open driven system connected to a thermal bath need not reach such an end.
Indeed, when the system and system-bath coupling Hamiltonians can be made time-independent through a simple rotating frame transformation, Gibbs-type steady states are expected~\cite{Galitskii1970, Iadecola2013, Shirai2014, Liu2014}.
More generically, however, even spontaneous emission into a zero temperature bath may cause heating due to the possibility of absorbing energy from the driving field.
How to control the steady states of driven systems, and, in particular, under what conditions they may be used to explore novel topological phenomena, remain a challenging problem.

Our aim in this work is to uncover new means to control the steady state occupations of Floquet-Bloch states in driven systems. Here we focus on the dissipative open-system dynamics governed by the system's coupling to external baths; the complicated problem of electron-electron interactions will be addressed in future work.
In particular we target the case of a half-filled fermionic system, where we seek to obtain an insulator-like steady state in which the lower Floquet band is filled and the upper Floquet band is empty.
We refer to this state as a Floquet insulator.
We investigate how this state can be approached through the relaxation of momentum and energy,
enabled by connection to low temperature bosonic and fermionic baths, see Fig.~\ref{fig:setup}.
For a semiconductor-based realization, these baths naturally correspond to phonons and the electromagnetic environment (bosonic baths), and to a lead connected to the system (fermionic bath); analogous couplings can be arranged, e.g., in cold atomic systems~\cite{Zwierlein2012, Grenier2014}.

Several dynamical processes and their interplay govern the density of particle-hole ``excitations'' above the ideal Floquet insulator state\cite{footnote:excitations}.
Radiative recombination constitutes an important mechanism for generating excitations.
In a non-driven system, recombination allows electrons in the conduction band to annihilate with holes in the valence band via the spontaneous emission of a photon. For resonant driving, as illustrated for the case of a one-dimensional system in Fig.~\ref{fig:setup}b, the Floquet bands feature a band inversion: 
states with crystal momenta between the two resonance values $\pm k_R$ in the {\it lower} Floquet band are primarily formed from {\it conduction band} states of the non-driven system,
while in the same interval the upper Floquet band is comprised of \textit{valence band} states. Therefore, radiative recombination results in transitions from the {\it lower} to the {\it upper} Floquet band, thus increasing the density of excitations, see Fig.~\ref{fig:setup}c.
Phonon scattering, on the other hand, enables relaxation of momentum and quasi-energy within and between Floquet bands, and may balance the recombination-induced heating.
A fermionic reservoir provides additional channels for removing excitations from the system, and also gives means to tune its total carrier density.
Importantly, photon-assisted electron-phonon scattering, as well as photon-assisted tunneling to/from the Fermi reservoir, generally also contribute to heating\cite{Kohler2005, Iadecola2014}, see Fig.~\ref{fig:filter}.

Our main message is that the driven electronic system can approach the Floquet insulator steady state when appropriately coupled to phonon and Fermi reservoirs.
In order for this to work, the coupling to the fermionic reservoir must be ``engineered'' to avoid the deleterious effects of photon-assisted tunneling. This can be accomplished by connecting the
system to the reservoir via a narrow-band energy filter (see Fig.~\ref{fig:filter} and Sec.~\ref{subsec:filtering}).  We also discuss regimes in which photon-assisted electron-phonon scattering can be suppressed. Most remarkably, at low temperatures and with energy-filtered coupling to a fermionic reservoir, we find that the driven system exhibits incompressible and insulating behavior. This implies that a steady-state Floquet topological insulator phase may be within reach.

\subsection{Structure of the paper and main results}

Before beginning the analysis, we briefly summarize the structure of the text to follow.
Keeping in mind our motivation of realizing Floquet topological insulators, our main focus in this work is on achieving Floquet insulator steady states.

First, in Sec.~\ref{sec:FloquetStates} we introduce the Floquet states of the periodically-driven lattice system, with band structure depicted schematically in Fig.~\ref{fig:setup}b.
After defining the Floquet states, we introduce the Floquet kinetic equation, Eq.~\eqref{eq:KineticEq}, which forms the basis for the description of many-body population dynamics throughout this work.
The kinetic equation can be obtained systematically from the exact (infinite) hierarchy of equations of motion for multi-particle correlators (see Appendix~\ref{appendix: kinetic}), and at our level of approximation takes on a simple intuitive form in terms of incoming and outgoing rates for each state.

Next, in Sec.~\ref{sec:bosons} we study the steady states when the system is coupled only to the bosonic baths.
Here our aim is to elucidate the competition between heating due to radiative recombination and momentum/energy relaxation by phonons in a particle number conserving system.
When the electron-phonon scattering rates (ignoring Pauli blocking) are large compared to the recombination rate, we find that the driven system approaches a Floquet insulator state, with separate particle and hole densities in the upper and lower Floquet bands, respectively, see Fig.~\ref{fig:Results1} below.
The steady state excitation density depends on the ratio of phonon-assisted inter-Floquet-band relaxation and recombination rates, becoming small for fast interband relaxation.
Using rather general arguments, we show that the steady state excitation density scales with the square root of the recombination rate in the limit of fast interband relaxation.
As a result, even strong electron-phonon coupling may be insufficient to fully deplete excited carriers from the system.

In Sec.~\ref{sec:lead} we introduce coupling to a fermionic reservoir. In Fig.~\ref{fig:Results2} we display the steady states for both wide-band and energy-filtered reservoirs. We show that coupling to a wide-band reservoir increases the density of excitations, due photon-assisted tunneling. The energy-filtered system-reservoir coupling suppresses all photon-assisted tunneling, and our results demonstrate that it can further \textit{reduce} the density of excitations.
We discuss two coupling geometries, where the Fermi reservoir is either coupled to the system at a single point (as a lead), or uniformly throughout the system.
For homogeneous coupling, when the chemical potential of the filtered reservoir is set inside the Floquet gap, the excitation density may be highly suppressed, thus bringing the system close to the ideal Floquet insulator state.
Interestingly, even when the steady state hosts a finite density of excitations, the system is incompressible in the sense that the excitation density is unaffected by small shifts of the chemical potential of the reservoir, see Fig.~\ref{fig:Results3}. For a lead coupled at a single point, the steady state distribution is necessarily inhomogeneous.
We provide an estimate for the ``healing length''  over which the distribution can be affected by such coupling.
Beyond this length, the steady state is set by the competition between recombination and electron-phonon coupling, as described in Sec.~\ref{sec:bosons}.

Finally, in Sec.~\ref{sec:discussion} we summarize the main results and discuss implications for transport experiments.
We discuss the corresponding observables and the conditions under which insulating behavior could be observed.

\section{Floquet-Bloch kinetic equation for the driven two-band system}
\label{sec:FloquetStates}

In this section we describe the single-particle properties of an isolated periodically-driven system.
We first give the Hamiltonian of the system without driving, and then discuss the form of driving and the resulting Floquet states.
We then introduce the kinetic equation for Floquet-state occupation numbers, which is the foundation for the description of many-body dynamics used throughout this work.
The section concludes with a brief overview of the dynamical processes described by the kinetic equation.

\subsection{System Hamiltonian and Floquet-Bloch states}

We now introduce the single-particle Hamiltonian and Floquet-Bloch states for the periodically-driven system that we consider.
Many of the features that we describe, including the form of the kinetic equation, hold quite generally, independent of dimensionality.
For concreteness, and to allow comparison with detailed numerical simulations, we focus on the case of a one-dimensional system with two bands.

The single particle Hamiltonian of the driven system is defined as follows.
We assume that the driving field is spatially uniform, thus maintaining the translational symmetry of the lattice.
In this case the crystal momentum $k$ is conserved.
For each $k$ the evolution within the corresponding $2 \times 2$ Bloch space is given by the Hamiltonian $H(k) = H_0(k) + V(t)$, with
\be
\label{eq:Hamiltonian} H_0(k) =  \tfrac12 E_k\, (\vec{d}_k\cdot\bm{\sigma}),\quad V(t) = \tfrac12 V_0\, (\vec{g}\cdot\bm{\sigma})\cos \Omega t,
\ee
where $\pm \frac12 E_k$ are the energies of the conduction and valence bands, $\vec{d}_k$ and $\vec{g}$ are unit vectors, $V_0$ and $\Omega$ are the amplitude and angular frequency of the drive, and $\bm{\sigma}$ is a vector of Pauli matrices acting in the two-component orbital space (in this work we ignore spin).
For now we leave the values of $\vec{d}_k$ and $\vec{g}$ unspecified, giving explicit forms when discussing numerical results below.

To understand the nature of the coupling induced by driving, we rotate to the basis of conduction and valence band states, i.e., to the basis which diagonalizes $H_0(k)$.
Specifically, the Bloch eigenstates in the conduction and valence bands of the non-driven system satisfy $H_0(k)\Ket{u_{kc}} = \frac12 E_k\Ket{u_{kc}}$ and $H_0(k)\Ket{u_{kv}} = -\frac12 E_k\Ket{u_{kv}}$.
The driving term $V(t)$ in Eq.~\eqref{eq:Hamiltonian} is expressed in the basis of lattice orbitals, and naturally does not depend on $k$.
However, after rotating to the basis of conduction and valence band states for each $k$, the driving picks up a non-trivial $k$-dependent matrix structure
\be
\tilde{H}_0(k) = \tfrac12 E_{k}\sigma_z,\quad \tilde{V}(k, t) = \tfrac12V_0 (\tilde{\vec{g}}_k\cdot\bm{\sigma})\cos \Omega t,
\label{eq: Hamiltonian2}
\ee
where tildes indicate operators in the basis of conduction and valence bands, and $\tilde{\vec{g}}_k = \tilde{g}_{k,\parallel}\hat{\vec{z}} + \tilde{\vec{g}}_{k,\perp}$
is a unit vector determined by the relative orientations of $\vec{d}_k$ and $\vec{g}$ in Eq.~\eqref{eq:Hamiltonian}, broken down to $z$ and $x-y$ components.

When the system is isolated, the Floquet-Bloch states $\{\Ket{\psi_{k\pm}(t)}\}$ provide a convenient basis for describing its evolution.
Each state $\Ket{\psi_{k\pm}(t)}$ can be expressed as a sum over harmonics:
\be
\label{eq:FloquetState}\Ket{\psi_{k\pm}(t)} = \sum_{n=-\infty}^\infty e^{-i(\mathcal{E}_{k\pm} + n\hbar\Omega) t/\hbar}\Ket{\phi_{k\pm}^n},
\ee
where $\mathcal{E}_{k\pm}$ is the quasi-energy of $\Ket{\psi_{k\pm}(t)}$ and $\{\Ket{\phi_{k\pm}^n}\}$ is a {\it non-normalized} (and over-complete) set of states found by Fourier transforming the time-dependent $2 \times 2$ Schr\"{o}dinger equation~\cite{Shirley1965,Sambe1973} in the Bloch space for crystal momentum $k$.
The quasi-energies $\{\mathcal{E}_{k\pm}\}$ and harmonics $\{\Ket{\phi_{k\pm}^n}\}$ in Eq.~\eqref{eq:FloquetState} are only uniquely defined up to the gauge freedom ${\mathcal{E}'}_{\!\!k\pm} = \mathcal{E}_{k\pm} + m\hbar\Omega$, $\Ket{{\phi}_{k\pm}^{\prime n}} = \Ket{\phi_{k\pm}^{n+m}}$.
Here we fix the gauge by choosing $\mathcal{E}_{k\pm}$ within a single Floquet-Brillouin zone centered around a specific {\it energy} $E_0$, $E_0 -\hbar\Omega/2 \le \mathcal{E}_{k\pm} < E_0 + \hbar\Omega/2$.

Before discussing many-body dynamics, a few comments on the nature of the Floquet bands are in order.
We are interested in the case where the driving field photon energy $\hbar \Omega$ exceeds the bandgap $E_{\rm gap}$ of the non-driven system, see Fig.~\ref{fig:setup}b.
In the Floquet picture, the leading-order influence of the driving can be understood by first shifting the valence band up by the photon energy $\hbar\Omega$.
After shifting, the bands become degenerate at the resonance points~\cite{footnote:resonancePoints} $\pm k_R$ in the Brillouin zone where $E_{k_R} = \hbar\Omega$.
Here the driving opens avoided crossings, resulting in a gap $\Delta_{k_R} \approx V_0|\tilde{\vec{g}}_{k_R,\perp}|$ between the two Floquet bands. The resulting band structure is depicted in Fig.~\ref{fig:setup}c.
We center the Floquet zone on these resonances in the conduction band, setting $E_0 =  \frac12 \hbar\Omega$.
Throughout we assume that the bandwidth is narrow enough such that the two-photon resonance condition is never satisfied~\cite{footnote:twoPhotonCondition}, i.e., $E_{k} < 2\hbar\Omega$ for all $k$.

As discussed in the introduction, the resonant driving introduces a {\it band inversion} in the Floquet bands.
Furthermore, near the resonant momenta $k_R$ the Floquet bands are strongly hybridized superpositions of conduction and valence band states.
These features of the Floquet bands have important consequences both for controlling band topology~\cite{BHZ, HasanKaneRMP} and for the nature of many-body dynamics in the system, as we describe below.

\subsection{The Floquet kinetic equation}

Below we use the Floquet basis of single-particle states to describe the many-body dynamics of the driven system when it is coupled to bosonic and fermionic baths.
In particular, we aim to characterize the steady states of the system in terms of the Floquet state occupation numbers $F_{k\alpha} = \avg{f^\dagger_{k\alpha}(t)f_{k\alpha}(t)}$, where $f^\dagger_{k\alpha}(t)$ creates an electron in the state $\Ket{\psi_{k\alpha}(t)}$ at time $t$, with $\alpha = \pm$.
Focusing on the dynamics for time scales much longer than the driving period, we develop a kinetic equation in the Floquet basis 
to describe the net rate of change of the population in the Floquet state $\Ket{\psi_{k\alpha}}$ due to electron-phonon scattering, radiative recombination, and tunneling to and from the fermionic reservoir:
\be
\label{eq:KineticEq} \dot{F}_{k\alpha}\, =\,  I^{\rm ph}_{k\alpha}(\{F\}) +I^{\rm rec}_{k\alpha}(\{F\}) + I^{\rm tun}_{k\alpha}(F_{k\alpha}).
\ee
Here the ``collision integrals'' $I^{\rm ph}$, $I^{\rm rec}$,  and $I^{\rm tun}$ describe electron-phonon scattering, recombination, and tunnel coupling to the reservoir, respectively, and $\{F\}$ indicates the set of occupation factors for all momentum and band index values.
The key processes associated with each of these terms are represented schematically in Fig.~\ref{fig:setup}c.

The derivation of Eq.~\eqref{eq:KineticEq} is rather technical, so here we briefly summarize the approach (for details, see Appendix~\ref{appendix: kinetic} and, e.g., Ref.~\onlinecite{Kira2012}).
We begin by considering the equations of motion for the single-particle correlators $\avg{f^\dagger_{k\alpha}(t) f_{k\alpha}(t)}$.
Coupling to the bath degrees of freedom generates an infinite hierarchy of equations of motion involving correlators of higher and higher order.
We focus on a homogeneous system, in the regime where coherences between different Floquet states can be neglected (see below).
Using a standard cluster-expansion approach, we systematically truncate the equation of motion hierarchy and obtain transition rates which coincide with those given by the ``Floquet Fermi's golden rule.''
Below we frame the discussion in terms of these golden-rule transition rates, which provide a clear intuitive picture for all terms contributing to Eq.~\eqref{eq:KineticEq}.
We will use the rates to build up the specific forms of the collision integrals $I^{\rm ph}$, $I^{\rm rec}$,  and $I^{\rm tun}$.

In describing the dynamics of the system, it is important to note that the occupation factors $F_{k\alpha}$ do not generally give a complete description of the steady state.
However, when transition rates associated with the system-bath interaction are small in comparison with the Floquet gap $\Delta_{k_R}$, off-diagonal correlations such as $\avg{f^\dagger_{k\alpha}(t)f_{k\alpha'}(t)}$ are suppressed in the steady state (see Appendix~\ref{appendix: kinetic}).
Crucially, even if the scattering rates are large when the system is far from the steady state, Pauli blocking in the steady state can strongly suppress the phase space for scattering.
Therefore Eq.~\eqref{eq:KineticEq} can provide a good description of the steady state, even in parameter regimes where it does not give a faithful description of the transient dynamics.
The requirement that the steady state scattering rates remain small compared with the Floquet gap $\Delta_{k_R}$ provides an important consistency check, which we apply to all numerical simulations discussed below.
Note that the cluster expansion approach provides a powerful framework that can be used to incorporate the roles of coherences and non-Markovian dynamics, going beyond the regime studied here\cite{futureWork}.

\section{Electron-phonon coupling and recombination}
\label{sec:bosons}

In this section we discuss the steady states of the electronic system which result from the competition between radiative recombination and coupling to the phonon bath.
Both processes arise from the coupling of electrons to a bosonic bath, comprised of photons in the former case and phonons in the latter.
Formally, the collision integrals $I^{\rm rec}$ and $I^{\rm ph}$ describing these processes are very similar.
However, it is important to understand that they act in competition.
During recombination, an electron transitions from the non-driven conduction band to the valence band, while emitting a photon.
In terms of the Floquet bands, this process {\it promotes} an electron from the lower to the upper Floquet band (see Fig.~\ref{fig:setup}), thereby heating the electronic system and increasing the total number of excitations.
On the other hand, electron-phonon scattering primarily relaxes excited electrons to the bottom of the upper Floquet band (and similarly relaxes holes to the top of the lower Floquet band),
and allows excited electrons to relax back to the lower Floquet band, thereby reducing the number of excitations.

Note that the electron-phonon interaction may also play an adverse role in the system: photon-assisted scattering processes may increase the number of excitations.  We show that these processes can be effectively eliminated under suitable conditions on the phononic spectrum and the form of the drive. 
Even when these processes are eliminated, radiative recombination remains as a source of heating in our model.

The competition between electron-phonon scattering and recombination determines the steady state of the system.
We show that these steady states feature Fermi seas of excited electrons and holes, with separate chemical potentials and a temperature equal to that of the phonon bath.
Below we first analyze the kinetic equation in the presence of a generic bosonic bath.
We then input the specific details needed to describe recombination and scattering by acoustic phonons, and analyze the resulting steady states.
Finally, through an analytical estimate we show that for fixed electron-phonon coupling the steady state excitation density grows with the square root of the radiative recombination rate.

\subsection{Collision integral for a generic bosonic bath}
\label{subsec:generalBoson}

The bosonic bath is described by the Hamiltonian $H_{\rm b} = \sum_{\vec{q}} \hbar\omega_{\vec{q}}\, b^\dagger_{\vec{q}}b_{\vec{q}}$, where $b^\dagger_{\vec{q}}$ and $b_{\vec{q}}$ are the creation and annihilation operators for a bosonic excitation carrying (crystal) momentum $\vec{q}$, and $\omega_{\vec{q}}$ is the corresponding frequency.
Using the creation and annihilation operators $\{c^\dagger_{k\nu}, c_{k\nu}\}$ for electrons in the bands of the non-driven system, defined below Eq.~\eqref{eq:Hamiltonian},
we describe the ``electron-boson'' interaction via $H_{\rm int} = \sum_{\vec{q}} H_{\rm int}(\vec{q})$, with
\be
\label{eq:ElectronBoson}H_{\rm int}(\vec{q}) = \sum_{k k'} \sum_{\nu,\nu'} G^{k'\nu'}_{k\nu}(\vec{q}) c^\dagger_{k',\nu'}c_{k\nu}(b^\dagger_{\vec{q}} + b_{-\vec{q}}).
\ee
Here $G^{k'\nu'}_{k\nu}(\vec{q})$ is the matrix element for scattering an electron with crystal momentum $k$ in band $\nu$ to crystal momentum $k'$ in band $\nu'$, with the emission (absorption) of a boson of momentum $\vec{q}$ ($-\vec{q}$).
We take the bath to be three dimensional, and the electronic system to lie along the $x$-axis.
Note that in Eq.~(\ref{eq:ElectronBoson}) we did not impose $k' = k - q_x$ to allow the possibility of describing a finite system coupled to a bath of larger size (such as the electromagnetic environment).
If the lengths of the bath and system are the same, we can impose conservation of the corresponding crystal momentum component, whereby $G^{k'\nu'}_{k\nu}(\vec{q})$ is nonzero only if $k'=k-q_x$.

As a fundamental building block for constructing the many-body collision rates, we calculate the rate $W^{k'\!\alpha'}_{k\alpha}$ for a {\it single electron} in an otherwise empty system to scatter from crystal momentum $k$ in Floquet band $\alpha$ to crystal momentum $k'$ in Floquet band $\alpha'$. For transparency, we focus on zero temperature; the analogous expressions at finite temperature are given in Appendix~\ref{appendix: kinetic}. For simplicity we take the matrix elements in Eq.~\eqref{eq:ElectronBoson} to depend only on $q_x$, i.e.,~$G^{k'\nu'}_{k\nu}(\vec{q}) = G^{k'\nu'}_{k\nu}(q_x)$; the discussion that follows can be easily generalized beyond this assumption, but the qualitative results will not be affected.

Due to the harmonic structure of the time-dependent Floquet state wave functions, Eq.~(\ref{eq:FloquetState}), the transition rate is given by a sum over many contributions, $W^{k'\!\alpha'}_{k\alpha} = \sum_n W^{k'\!\alpha'}_{k\alpha}\!(n)$.
In terms of the electronic operator $\hat{G}(q_x) \equiv \sum_{k,k'}\sum_{\nu,\nu'} G^{k'\nu'}_{k\nu}(q_x) c^\dagger_{k',\nu}c_{k\nu'}$, these contributions are given by
\be
\label{eq:FFGR}W^{k'\!\alpha'}_{k\alpha}\!(n) = \frac{2\pi}{\hbar} \sum\limits_{q_x} \Big\vert\sum_m\MatEl{\phi_{k'\alpha'}^{m+n}}{\hat{G}(q_x)}{\phi_{k\alpha}^{m}}\Big\vert^2 \rho_{q_x}(-\Delta \mathcal{E}_n),
\ee
where $\Delta\mathcal{E}_n = \mathcal{E}_{k'\alpha'} \mathcal - \mathcal{E}_{k\alpha} + n\hbar\Omega$. The quasi-energy difference between final and initial electronic states is $\mathcal{E}_{k'\alpha'} \mathcal - \mathcal{E}_{k\alpha}$, and
$\rho_{q_x}(\omega)$ is the boson density of states at frequency $\omega$ for a fixed value of the boson's longitudinal momentum component $q_x$.
Note that for a monotonic boson dispersion, $\rho_{q_x}(\omega)$ is only nonzero if $\omega > \omega_{\vec{q}_0}$, where $\vec{q}_0 = (q_x, 0, 0)$.
The scaling of the individual rates $W^{k'\!\alpha'}_{k\alpha}$ with system size is discussed in Appendix \ref{appendix: scaling}.

\begin{figure}[t]
\begin{center}
\includegraphics[width=1.0\columnwidth]{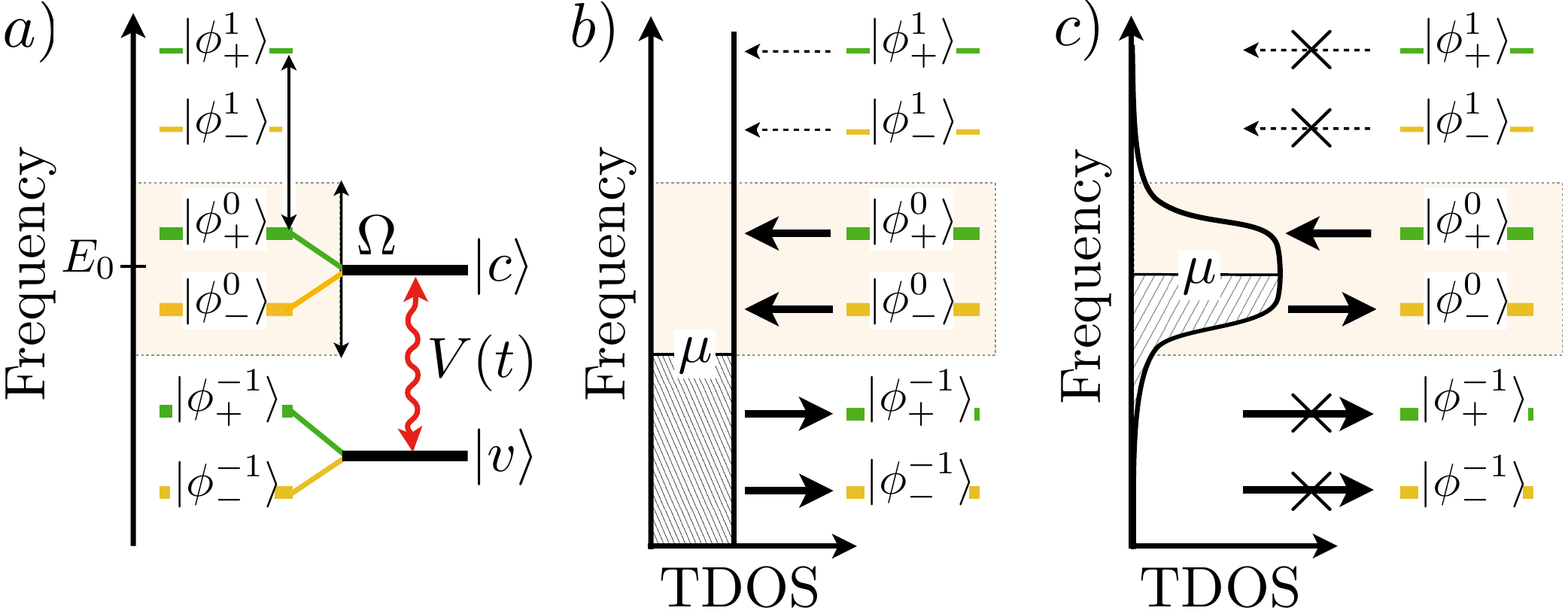}
\caption{Harmonic structure of Floquet states and energy-filtered reservoir coupling.
a) Floquet harmonics of a two-level system with states $\Ket{v}$ and $\Ket{c}$ coupled by an {\it on-resonance} driving field $V(t)$.
The Floquet zone (shaded) is centered at the energy $E_0$, set equal to the energy of the resonant state $\Ket{c}$.
In the special case of a rotating-field, $V(t) = \tfrac12 V_0e^{-i\Omega t}\Ket{c}\Bra{v} + {\rm h.c.}$, we have $\Ket{\phi_\pm^0} = \Ket{c}$, $\Ket{\phi_\pm^{-1}} = \pm \Ket{v}$ and $\mathcal{E}_\pm = E_0 \pm \tfrac12 V_0$, see Eq.~\eqref{eq:FloquetState}.
Away from resonance, the relative normalizations of $\Ket{\phi_+^{n}}$ and $\Ket{\phi_-^{n}}$ will change.
For a more general form of weak driving, the dominant harmonics are shown in bold.
b) The Floquet states $\Ket{\psi_\pm(t)}$ are both coupled to filled and empty states of a wide-band reservoir via the harmonics $\{\Ket{\phi_\pm^n}\}$, see Eq.~\eqref{eq:FFGR_Lead}.
Here the reservoir chemical potential is set in the gap of the non-driven system.
c) When coupling is mediated by a narrow-band energy filter, the tunneling density of states (TDOS) and photon-assisted tunneling are suppressed outside the filter window.
By setting the reservoir chemical potential inside the Floquet gap, centered around the energy $E_0$ in the original conduction band (see Fig.~\ref{fig:setup}b), the lower and upper Floquet bands are selectively filled and emptied, respectively.
}
\label{fig:filter}
\end{center}
\end{figure}

The structure of the transition rates in Eq.~\eqref{eq:FFGR} can be understood heuristically as follows.
Superficially, $\Ket{\psi_{k\pm}(t)}$ in Eq.~\eqref{eq:FloquetState} takes the form of a superposition over a ladder of states $\Ket{\phi^n_{k\pm}}$ with ``energies'' $\mathcal{E}_{k\pm} + n\hbar\Omega$, see illustration in Fig.~\ref{fig:filter}a.
Viewing these harmonics $\Ket{\phi^n_{k\pm}}$ as independent states, the net transition rate $W^{k'\!\alpha'}_{k\alpha} = \sum_n W^{k'\!\alpha'}_{k\alpha}\!(n)$ is found by summing the contributions from all pairs of initial and final states, while taking into account ``energy'' conservation.
The appearance of $n\hbar\Omega$ inside the density of states in Eq.~\eqref{eq:FFGR} expresses the fact that quasi-energy is a periodic variable, and therefore Floquet scattering processes need only conserve quasi-energy \emph{up to multiples of the driving field photon energy} $\hbar\Omega$. If a boson is emitted and an electron {\it decreases} its quasi-energy, $\Delta\mathcal{E} < 0$, then the scattering rate can be non-zero for $n = 0$. Interestingly, the scattering rate can also be non-zero
if a boson is emitted and an electron {\it increases} its quasi-energy, $\Delta\mathcal{E} > 0$, if $n < 0$.

The collision integrals in Eq.~\eqref{eq:KineticEq} are given by the differences between the total rates for scattering into and out of the state $\Ket{\psi_{k\alpha}(t)}$, due to recombination or coupling to acoustic phonons.
In turn, these rates are obtained by multiplying the bare rates in Eq.~\eqref{eq:FFGR} by products of occupation factors $F_{k\alpha}$, $\bar{F}_{k'\alpha'} \equiv (1 - F_{k'\alpha'})$, etc., to account for the filling of the initial and final states:
\be
\label{eq:CollisionInt} I_{k\alpha} = \sum_{k'\alpha'}\left[W^{k\alpha}_{k'\alpha'}\bar{F}_{k\alpha}F_{k'\alpha'} - W^{k'\!\alpha'}_{k\alpha} \bar{F}_{k'\alpha'}F_{k\alpha}\right].
\ee
The corresponding expressions for nonzero bath temperature are shown in the Appendix~\ref{appendix: kinetic}.

Equations~\eqref{eq:FFGR} and \eqref{eq:CollisionInt} support what we refer to as ``Floquet-Umklapp'' processes, in which quasi-energy conservation is satisfied with $n \neq 0$. Such processes generically heat the system when they are allowed within the kinematic constraints imposed by the bath and Floquet-system spectra (i.e.~by energy and momentum conservation). For example, even at zero bath temperature, an electron may spontaneously jump from the {\it lower} Floquet band to the {\it upper} one while {\it emitting} a bosonic excitation
(see Fig.~\ref{fig:setup}c). As we show below, such processes cause deviations from the ideal Floquet insulator state.

Fortunately, Floquet-Umklapp processes are suppressed under appropriate conditions on the dispersion of the bath bosons.
In fact, Floquet-Umklapp processes are completely suppressed if the bath bandwidth is limited such that $W^{k'\!\alpha'}_{k\alpha}\!(n)$ strictly vanishes for all $n \neq 0$.
Practically speaking, this means that the maximal boson energy $\hbar\omega^{\rm max}_{\vec{q}}$ must be smaller than the quasi-energy gap at the Floquet zone edge (i.e.,~the gap around $\hbar\Omega/2$ in Fig.~\ref{fig:setup}c), such that the energy conservation condition $-\Delta\mathcal{E} - \hbar\omega^{\rm max}_{\vec{q}} - n\hbar\Omega = 0$ cannot be satisfied with $n \neq 0$.
Below we will show how Floquet-Umklapp processes are manifested in radiative recombination and phonon scattering processes, and discuss methods to suppress them.

\subsection{Radiative recombination}
\label{subsec:radiative}

Having established the general framework for coupling the driven system to a bosonic bath, we now use it to study specific dissipation mechanisms which are relevant for driven semiconductor systems. 
We start by considering radiative recombination.

In non-driven systems, radiative recombination occurs when an excited particle in the conduction band relaxes to fill a hole in the valence band.
This results from the interaction of electrons with the electromagnetic environment, which is represented by a bosonic bath in our model.
In typical semiconductors, the electromagnetic interaction leads to transitions between states of {\it different} bands.
This restriction on the transitions arises due to two important factors: 1) the large speed of light implies that energy and momentum conserving transitions are practically ``vertical'' (i.e.~the electronic momentum is conserved), and 2) the electromagnetic dipole matrix elements couple states from {\it different} atomic orbitals.
To impose this restriction in our model we describe the interaction with the electromagnetic environment using matrix elements of the form $G^{\rm rec} \propto (1 - \delta_{\nu\nu'})$. For simplicity, in the simulations below we model vertical recombination via~\cite{footnote:recombination} $G^{\rm rec}  = g^{\rm rec}(1-\delta_{\nu\nu'})\delta_{q_x,0}\delta_{k,k'}$, and take a constant density of states $\rho_0$  for photons with energies $\hbar\omega \gtrsim E_{\textrm{gap}}$.

We now describe the processes resulting from the coupling to the electromagnetic environment in the {\it driven} system that we consider.
Following from the situation in the non-driven case, relaxation via emission of a photon to the environment is possible from a Floquet state of predominantly conduction band character to one of predominantly valence-band character.
Due to the band inversion described in detail in Sec.~\ref{sec:FloquetStates}, the $-$ Floquet band has conduction band character for momenta $|k| < k_R$.
Therefore, spontaneous transitions from the $-$ to the $+$ Floquet band are possible for states in this momentum range. 
Note that these Floquet-Umklapp processes {\it increase} the total electronic quasi-energy, and play an important role in determining the density of excitations in the steady state of the system (see Sec.~\ref{subsec: bosons steady}).
The rates of these processes may be controlled to some extent by placing the system in a cavity or photonic crystal, which modifies the photon density of states.
In addition, spontaneous transitions from the $+$ to the $-$ Floquet band are allowed in the momentum region $|k| > k_R$, where the $+$ Floquet band has predominantly conduction band character.
These processes help to reduce the total electronic quasi-energy, but will play a minor role near the steady state where the $+$ Floquet band is mostly empty.

The processes we have considered so far follow directly from those that are active in a non-driven system.
However, in a driven system an electron may also transition from a state of valence band character to one of conduction band character, by  emitting a photon to the environment while absorbing energy from the drive.
Such processes are possible for initial states in the $-$ Floquet band with $|k| > k_R$, and for initial states in the $+$ Floquet band with $|k| < k_R$.
Referring to Eq.~(\ref{eq:FFGR}), the matrix elements for these processes\cite{footnote:smallPhoton} are suppressed by $[V_0/(\hbar\Omega)]^2$ for weak driving, and hence their rates are suppressed as $[V_0/(\hbar\Omega)]^4$.

\subsection{Scattering due to acoustic phonons}
\label{subsec: phonons}

The interaction between the electronic system and a bath of acoustic phonons plays a key role in setting the steady state of the driven system.
 Phonon-mediated scattering quickly relaxes excited electrons (holes) to the bottom (top) of the respective Floquet band.
In addition, phonon-mediated scattering allows these excitations to relax across the Floquet gap.
The competition between the latter interband scattering processes and radiative recombination sets the steady state density of excitations, as we discuss below.

In our model we assume that the electron-phonon coupling conserves the band index $\nu$ of the non-driven system, $G^{\textrm{ph}} \propto \delta_{\nu\nu'}$, as is typical for wide gap semiconductors \cite{Shah99,Yu10, Kira2012}.
The coherent drive hybridizes the bands near the resonances $\pm k_R$, thus enabling both intraband {\it and} interband scattering in the Floquet bands (see Fig.~\ref{fig:setup}c).
Note that the scattering crucially involves the exchange of both crystal momentum and quasi-energy between the phonons and the electrons, thus allowing relaxation of these quantities. We take the matrix elements  to conserve lattice momentum, $G^{k'\nu'}_{k\nu}(q_x)  = g(q_x)\delta_{\nu\nu'}\delta_{q_x,k-k'}$. In principle, the $q_x$ dependence of $g(q_x)$ depends on the specific type of electron-phonon coupling. For simplicity, we take the matrix elements to be independent of $q_x$, but have numerically verified that other choices do not change the qualitative results.

Besides helping to relax excitations, photon-assisted electron-phonon scattering can increase the excitation density.
Such Floquet-Umklapp scattering transfers electrons from the lower to the upper Floquet band, and can occur even for a zero temperature phonon bath.

Phonon-related Floquet-Umklapp processes can be suppressed in several ways.
First, as discussed in Sec.~\ref{subsec:generalBoson}, limiting the bandwidth for the phonons to be smaller than the quasi-energy gap at the Floquet zone edge, $\Delta_{\textrm{edge}}$, efficiently suppresses photon-assisted scattering.
Note, however, that the phonon bandwidth should remain bigger than the Floquet gap emerging at the resonance momenta, $\Delta_{k_R}$, as otherwise phonons would be unable to facilitate relaxation between the upper and lower Floquet bands. An optimal phonon bandwidth $\omega_D$ would therefore satisfy $\Delta_{k_R}<\hbar \omega_D<\Delta_{\textrm{edge}}$.
The bandwidth for the phonon bath depends on material parameters, however, and may not be easily tunable.

Interestingly, additional routes are available for suppressing Floquet-Umklapp processes involving phonons. If the boson bandwidth allows the energy conservation condition $-\Delta\mathcal{E} - \hbar\omega^{\rm max}_{\vec{q}} - n\hbar\Omega = 0$ to be satisfied for $|n| \le 1$ (but not for $|n| > 1$), the rates $W^{k'\!\alpha'}_{k\alpha}\!(n)$ with $n = \pm 1$ can be controlled by the choice of driving.
In particular, for harmonic driving they vanish as $\tilde{g}_\parallel \rightarrow 0$ (see Appendix~\ref{appendix: kinetic}).
For many experimentally-relevant materials driven by optical fields, $\tilde{g}_\parallel$ is indeed small for momenta near $k = 0$.
Additionally, even when none of the conditions above are met, the amplitudes of the Floquet harmonics $\{\Ket{\phi_{k\alpha}^{n}}\}$ (and hence the rates $W^{k'\!\alpha'}_{k\alpha}\!(n)$) are generically suppressed for large $n$.
Hence, although heating inevitably accompanies coupling to a bosonic bath, there are many ways to control or limit the corresponding effects on the steady state distribution
(see below and also Refs.~\onlinecite{Galitskii1970, Iadecola2013, Shirai2014, Liu2014}).

\subsection{Steady state}
\label{subsec: bosons steady}

\begin{table}
\centering

\begin{tabular}{|c|c|c|c|c|}
\hline
$A$ & $E_{\textrm{gap}}$ & $\mathbf{g}$ & $\mathbf{d}_\mathbf{k}$ & $V_{0}$\tabularnewline
\hline
$0.25\hbar\Omega$ & $0.8\hbar\Omega$ & $(1,0,0)$ & $(0,0,1)$ & $0.1\hbar\Omega$\tabularnewline
\hline
\hline
$c_{s}$ & $\hbar\omega_{D}$ & \multicolumn{2}{c|}{$2\pi(G_{0}^{ph})^{2}\bar{\rho}^{\rm{ph}}$} & $k_{B}T$ \tabularnewline
\hline
$\frac{0.05}{\pi\sqrt{3}}a\Omega$ & $0.15\hbar\Omega$ & \multicolumn{2}{c|}{$\p{2\times10^{-2}}\hbar\Omega$} & $0.1\Delta_{k_{R}}$  \tabularnewline
\hline
\end{tabular}
\caption{
Parameters fixed in all simulations. Top row: parameters of the electronic Hamiltonian, Eq.~(\ref{eq:Hamiltonian}), with $E_{k}=2A[1-\cos(ka)]+E_{\textrm{gap}}$, where $a$ is the lattice constant. The drive is spatially uniform, $V(t)=\frac12 V_0 (\mathbf{g}\cdot \bm{\sigma})\cos\Omega t$. Bottom row: parameters of the three dimensional acoustic phonon bath, where $c_s$ is the phonon velocity, and $\omega_{D}$ is the Debye frequency. In all simulations, the overall scale of the phonon matrix elements is set by fixing the ratio $2\pi(G_{0}^{ph})^{2}\bar{\rho}^{\rm{ph}}/(\hbar\Omega)$, where $\bar{\rho}^{\rm{ph}}$ is the phonon density of states at zero momentum and energy $\hbar c_{s}(\pi/a)$.
For convergence, in the simulations we keep the phonon bath at a small temperature, $k_{B}T\approx10^{-2}\hbar\Omega$.
}
\vspace{-0.2 in}
\label{table:parameters}
\end{table}

\begin{figure}[t]
\begin{center}
\includegraphics[width=1\columnwidth]{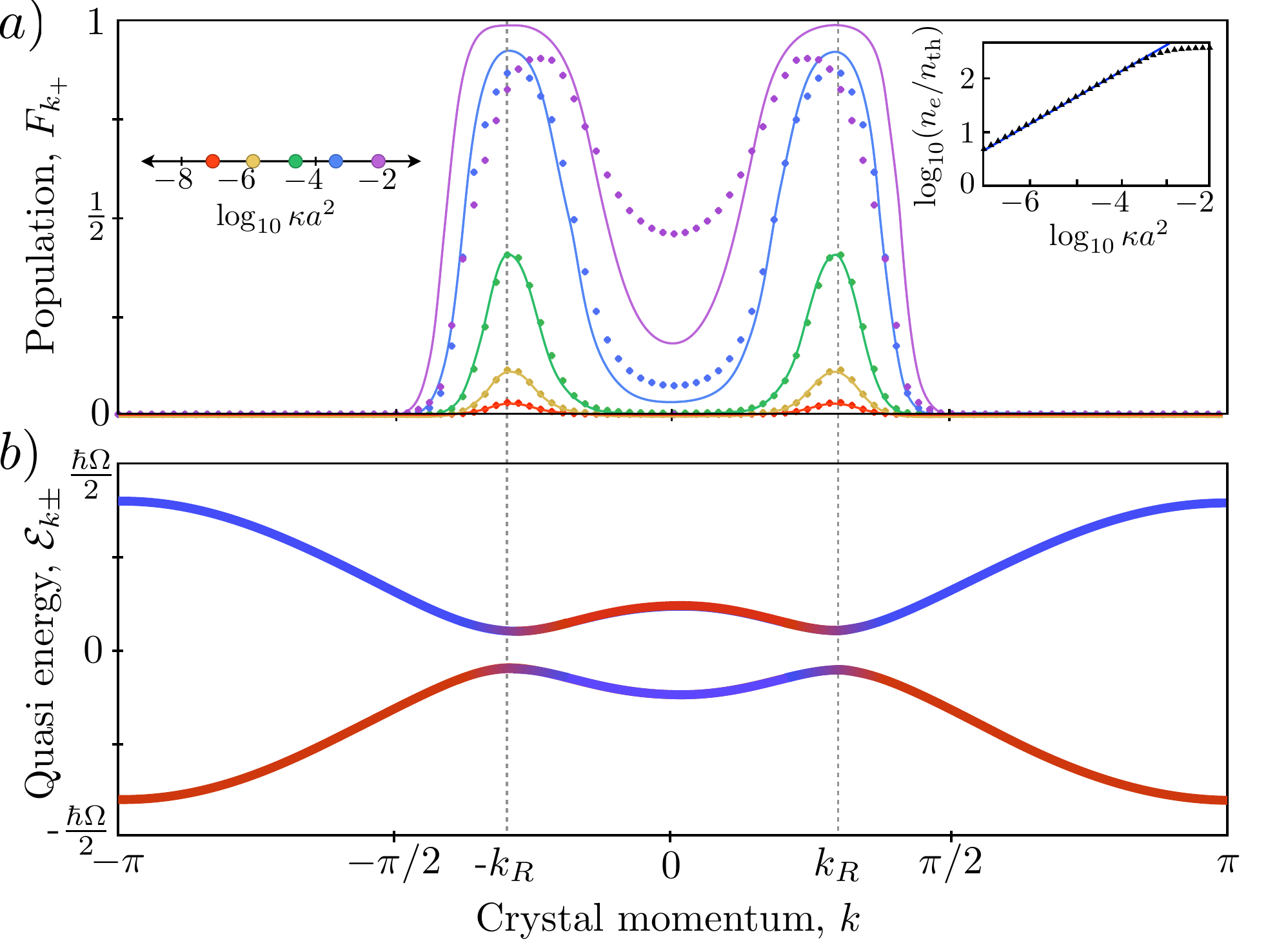}
\caption{Numerically obtained steady states with radiative recombination and coupling to acoustic phonons. 
Here the density is set to half-filling, and we use a 3D acoustic phonon bath with $\hbar\omega_D$ smaller than the gap $\Delta_{\rm edge}$ at the Floquet zone edge. The phonon temperature is set to $k_BT=10^{-2}\hbar\Omega$. We keep the phonon and photon densities of states fixed,
and only vary an overall scale for the coupling matrix elements.
The full details of the model can be found in Table~\ref{table:parameters}.
(a) Distribution of electrons in the upper Floquet band, $F_{k+}=\langle f^{\dagger}_{k+} f_{k+}\rangle$, for several values of
$\kappa=k_R\overline{\mathcal{W}}^{\rm rec}/\pi\Lambda^{\rm inter}$, see Eq.~(\ref{eq:rho_steady}) and Appendix~\ref{appendix: parameters} for definitions.
The distributions are fitted to a Floquet-Fermi-Dirac distribution at temperature $T$ (solid lines).
Due to particle-hole symmetry, the distributions of holes in the lower Floquet band, $1-F_{k-}$, are identical to the distributions shown.
{\it Inset}: Log-Log plot showing the total density of electrons in the upper Floquet band, $n_e$ as a function of $\kappa$. The density $n_e$ is normalized to the ``thermal density'' $n_{\textrm{th}}=6.8\times 10^{-4}$ (see text).
The plot demonstrates the square root behavior predicted in Eq.~\eqref{eq:rho_steady}.
Note that for large $n_e$, the recombination rates saturate due to Pauli blocking.
The Floquet band structure is shown in panel (b).}
\label{fig:Results1}
\end{center}
\vspace{-0.2 in}
\end{figure}

The steady state of the driven model described above results from the competition between the two main dissipation mechanisms: radiative recombination and acoustic phonon scattering.
To gain a more quantitative picture of the behavior, we numerically solve for the steady states of the kinetic equation~\eqref{eq:KineticEq} in the model outlined above, with the parameter values given in Table~\ref{table:parameters}. We take acoustic phonons to have a linear dispersion in three dimensions, $\omega^{\rm ph}_{\vec{q}} = c_s|\vec{q}|$, up to a ``Debye frequency'' cutoff $\omega_D$. We focus on the situation where $\Delta_{k_R}<\hbar\omega_D<\Delta_{\textrm{edge}}$, which allows inter-Floquet-band scattering, but forbids Floquet-Umklapp phonon scattering processes.
The rates $\{W^{k\alpha}_{k'\alpha'}\}$ are calculated using the form for the matrix elements described in Sections~\ref{subsec:radiative}~and~\ref{subsec: phonons}. 
Our results are summarized in Fig.~\ref{fig:Results1}.

The main result of the numerical simulation is the steady state distribution $F_{k+}$ of excited carriers in the upper Floquet band, which is shown in Fig.~\ref{fig:Results1}a, for a total density of particles corresponding to half-filling.
Due to particle-hole symmetry, the distribution of holes in the lower Floquet band, $\bar{F}_{k-}$, is identical to that above.
We examine the behavior of the steady state distribution while tuning the ratio of the phonon scattering and radiative recombination rates.
Specifically, we fix the parameters for electron-phonon coupling, and vary the overall scale of the matrix elements for recombination.

As seen in Fig.~\ref{fig:Results1}a, in all cases the upper Floquet band hosts a finite density of excited Floquet carriers, localized around the two band minima.
For relatively weak electron-phonon coupling the excitation density is large, but is limited by saturation of the recombination rates due to Pauli exclusion above a given density.
Notably, when electron-phonon coupling is relatively strong, the excitation density is significantly suppressed.
Moreover, in this regime the distribution of excited carriers is well described by a Floquet-Fermi-Dirac distribution with an effective chemical potential $\mu_e$ (a fit parameter) and temperature corresponding to that of the phonon bath (solid lines).
By ``Floquet-Fermi-Dirac distribution'' we refer to a distribution of particles in Floquet states, which is described by a Fermi-Dirac distribution taken as a function of quasi-energy.
The distribution of holes in the lower Floquet band (not shown) takes an identical form with an effective chemical potential $\mu_h = \mu_e$ due to particle-hole symmetry of the model.
To check the consistency of our approach, we verify that the scattering rates in the steady state are significantly smaller than the Floquet gap $\Delta_{k_R}$.
This condition is satisfied in particular for momenta around $k_R$ where the electron and hole excitation densities are localized.
A more detailed discussion of the scattering rates is provided in Appendix \ref{appendix: rates}.

The above form for $F_{k+}$ can be understood by considering the dynamics of the electrons coupled to the photon and phonon baths. When an electron is excited to the upper Floquet band via a recombination process, it quickly ``trickles down'' via repeated intraband scattering from acoustic phonons until eventually reaching one of the minima of the band.
There it joins the Fermi gas of excited electrons.
Relaxation from the upper to the lower band via phonon emission is only substantial near the band bottom, where the original valence and conduction bands are strongly hybridized.
The total density of excited carriers is determined by a balance between the interband excitation and relaxation process.

As seen in Fig.~\ref{fig:Results1}a, even for relatively large electron-phonon coupling strengths the density of excited electrons remains appreciable.
As we now explain, this situation arises from a bottleneck in interband relaxation due to the suppression of phonon emission rates for small excitation densities.

The relaxation bottleneck can be understood by considering the rate of change of the excitation density $n_e = \int \frac{dk}{2\pi} F_{k+}$ of excited electrons in the upper Floquet band.
In a heuristic model for the regime of low excitation density, recombination transfers electrons from the mostly filled states in the ``valley''  between maxima of the lower Floquet band (centered around $k = 0$), to the mostly-empty ``hump'' in the upper band, providing a constant source term for the excitations (see Fig.~\ref{fig:setup}c): $\dot{n}_e^{\rm rec} = \gamma^{\rm rec}$, with
\be
\label{eq:gamma_rec} \gamma^{\rm rec}  \approx \int_{-k_R}^{k_R} \frac{dk}{2\pi}\,\mathcal{W}^{\rm rec}_{k}.
\ee
Here, $ \mathcal{W}^{\rm rec}_{k} \equiv \sum_{k'} W_{k-}^{k'+}$
is the total rate for an electron, initially in the lower Floquet band with momentum $k$, to ``decay'' to the upper Floquet band with {\it any} final momentum (within the constraints of quasi-energy and crystal momentum conservation).
Thus $\mathcal{W}^{\rm rec}_{k}$ is simply the recombination rate for a single electron.
When we compute $\gamma^{\textrm{rec}}$ using $\mathcal{W}^{\rm rec}_k$, we take the occupations in the lower and upper Floquet bands in the interval $-k_R \le k \le k_R$ to be one and zero, respectively.
It is convenient to define an average recombination rate in this interval, $\overline{\mathcal{W}}^{\rm rec}$, whereby Eq.~(\ref{eq:gamma_rec}) becomes $\gamma^{\rm rec} \approx  (k_R/\pi)\overline{\mathcal{W}}^{\rm rec}$.

Relaxation via \textit{interband} electron-phonon scattering occurs for momenta in narrow regions around $\pm k_R$, from the bottom of the upper Floquet band to the top of the lower Floquet band.
For simplicity, in the discussion below we set $W_{k+}^{k'-} = \overline{W}^{\rm inter}$, where $\overline{W}^{\rm inter}$ is an average value for the transition rates in the active regions around $\pm k_R$.
The {\it total rate} of electrons relaxing from the upper to the lower Floquet band is found by summing the transition rates from occupied states in the upper band to empty states in the lower band.
The corresponding change to the excitation {\it density} goes as $\dot{n}_e^{\rm inter} \approx \tfrac{1}{L}\sum_k\sum_{k'} \overline{W}^{\rm inter} F_{k+}\bar{F}_{k'-}$.
Using particle-hole symmetry of the distribution, and $\sum_k F_{k+} = L n_e$, we obtain $\dot{n}_e^{\rm inter} \approx  -\Lambda^{\rm inter}n_e^2$, where $\Lambda^{\rm inter} \equiv L\overline{W}^{\rm inter}$.

The two powers of excitation density appearing in the expression for $\dot{n}_e$ come from 1) the density of excited electrons available to decay and 2) the density of final states available for each electron.
When the phonon bath is at finite temperature, the picture above is valid when $n_e$ exceeds the thermal excitation density (see below).

Importantly, despite the system size $L$ appearing explicitly in the definition of $\Lambda^{\rm inter}$, the net relaxation rate is in fact {\it system size independent}.
As explained in Appendix~\ref{appendix: scaling}, the individual rates $W_{k\alpha}^{k'\alpha'}$ to scatter between specific momentum values $k$ and $k'$ generically scale as $1/L$.
The system size independence is restored by the increasing number of final states, which scales as $L$.

Combining the recombination and interband phonon scattering terms, we obtain $\dot{n}_e = \gamma^{\rm rec} - \Lambda^{\rm inter}n_e^2$.
The condition $\dot{n}_e = 0$ yields a relation for the steady state excitation density $n_{\rm steady}$:
\be
\label{eq:rho_steady} n_{\rm steady} \approx \left(\frac{k_R}{\pi}\frac{\overline{\mathcal{W}}^{\rm rec}}{{\Lambda}^{\rm inter}}\right)^{1/2}.
\ee
The square root dependence in Eq.~\eqref{eq:rho_steady} is clearly exhibited in our simulations\cite{footnote:sqrt}, as shown in the inset of Fig.~\ref{fig:Results1}a.

Note that in our simulations the bath temperature was set to $k_BT=0.1\Delta_{k_R}$.  
At this temperature, a ``global'' Floquet-Fermi-Dirac distribution  with its chemical potential set in the middle of the Floquet gap would have a small density of excited electrons, $n_{\textrm{th}}$ (and similarly for holes). 
Here we define the ``global'' Floquet-Fermi-Dirac distribution as a single distribution describing the electronic occupations in both bands of the system.
For very low recombination rates, the square root behavior should saturate when $n_e \approx n_{\textrm{th}}$. However, throughout the parameter range used for Fig.~\ref{fig:Results1}a,  $n_e\gg n_{\textrm{th}}$, and therefore the effect of the  finite temperature of the bath on the square root behavior is negligible.

To summarize this section, when radiative recombination and other Floquet-Umklapp processes are absent, the system approaches the ideal Floquet insulator state (at half filling).
Importantly, our analysis shows that Floquet-Umklapp processes cannot be ignored: the steady state excitation density rises rapidly when the recombination rate is increased from zero.
In order to further reduce the excitation density, additional controls are needed.
Coupling the system to a Fermi reservoir can provide such a control, which we shall study in detail in the next section.

\section{Coupling to a Fermi reservoir}
\label{sec:lead}

In this section we consider the steady state of the system upon coupling it to an external fermionic reservoir. Our motivation here is twofold: the reservoir serves as an additional effective control over the steady state of the system, and is a necessary component of transport experiments.
However, as we show below, when the driven system is coupled to a standard fermionic reservoir with a wide bandwidth, photon assisted tunneling significantly increases the density of excitations. In addition, even in the ballistic regime, photon assisted tunneling opens extra channels for transport \cite{Kohler2005,Gu11,Kitagawa2011,Kundu2013,Kundu2014}. We will discuss how such processes can be suppressed using energy filtering, thereby allowing for the possibility to control and probe the driven system using external fermionic reservoirs.

In the discussion below we first assume that the distribution remains homogeneous under coupling to the reservoir.
This can be approximately satisfied for small systems with point-like coupling to a lead, or for systems where the coupling is extended rather than pointlike.
Next we focus on the scenario of a lead coupled at a point, where we will discuss the role of inhomogeneities and the length scale over which the steady state distribution is controlled by the lead.

\subsection{Collision integral for a fermionic reservoir}
\label{subsec: I fermi}

The Hamiltonian of the isolated reservoir is given by $H_{\rm res} = \sum_\ell E_\ell\, d^\dagger_\ell d_\ell$, where $d^\dagger_\ell$ creates an electron in state $\Ket{\ell}$ of the reservoir with energy $E_\ell$.
Throughout this work we assume that the periodic driving acts only on the system, and does {\it not} affect the reservoir.
We describe tunneling between the reservoir and states of the (undriven) system by the Hamiltonian $H_{\rm tun} = \sum_{\ell, k\nu} J_{\ell,k\nu}\, (d^\dagger_\ell c_{k\nu} + c^\dagger_{k\nu}d_{\ell})$.
The values of the tunneling matrix elements $J_{\ell, k\alpha}$ depend on the precise forms of the reservoir states $\{\Ket{\ell}\}$, the Bloch wave functions of the undriven system, and the details of the coupling.

The Floquet states $\Ket{\psi_{k\pm}}$ are coupled to the Fermi reservoir via the harmonics $\Ket{\phi_{k\pm}^n}$, as shown in Fig.~\ref{fig:filter}b,c.  These harmonics are spread over a large range of frequencies $\mathcal{E}_{k\pm}+n\hbar\Omega$. As a result, both the upper and lower Floquet bands are coupled to reservoir states in a wide range of energies.
This coupling is directly mirrored in the collision integral for the reservoir. Following the spirit of the discussion surrounding Eq.~\eqref{eq:FFGR}, we define the ``bare'' rate $\Gamma^{n}_{k\alpha}$
for a single electron to tunnel from a (filled) reservoir into
the Floquet state $\Ket{\psi_{k\alpha}(t)}$, via the harmonic $|\phi_{k\alpha}^n\rangle$,
\be
\label{eq:FFGR_Lead}\Gamma^{n}_{k\alpha} = \frac{2\pi}{\hbar}\sum_{\ell} |\MatEl{\phi_{k\alpha}^n}{H_{\rm tun}}{\ell}|^2\delta(\mathcal{E}_{k\alpha} + n\hbar\Omega - E_\ell).
\ee
Next, we assume that the reservoir is in equilibrium, with the occupation of a state with energy $E_\ell = \mathcal{E}_{k\alpha} + n\hbar\Omega$ 
given by the Fermi-Dirac distribution $D(E_\ell)$
with chemical potential $\mu_{\rm res}$ and temperature $T_{\rm res}$.
To build up the integral $I_{k\alpha}^{\rm tun}$ in the kinetic equation~\eqref{eq:KineticEq}, we supplement the rates $\{\Gamma^n_{k\alpha}\}$ in Eq.~\eqref{eq:FFGR_Lead} with the occupation factors ${F}_{k\alpha}$ and $D({\mathcal{E}^{n}_{k\alpha}})$, with $\mathcal{E}^{n}_{k\alpha} \equiv \mathcal{E}_{k\alpha} + n\hbar\Omega$:
\be
\label{eq:I_Lead}I_{k\alpha}^{\rm tun} = \sum_n \Gamma^{n}_{k\alpha}\!\left[\bar{F}_{k\alpha}D({\mathcal{E}^{n}_{k\alpha}) - {F}_{k\alpha}\bar{D}(\mathcal{E}^{n}_{k\alpha}})\right].
\ee
The first and second terms of Eq.~(\ref{eq:I_Lead}) correspond to electrons tunneling into and out of the system, respectively.

\subsection{Steady state with fermionic and bosonic baths}
\label{subsec:steadyStateLead}

How does the coupling to the reservoir influence the steady state of the system?
The possibility of photon-assisted tunneling of particles between the system and the reservoir makes the behavior of the driven system strikingly different from its equilibrium behavior.

To understand the effect of the reservoir, it is instructive to first look at the steady state distribution $\tilde{F}_{k\alpha}$ of Eq.~\eqref{eq:KineticEq} in the absence of recombination and electron-phonon scattering, $I^{\rm rec} = I^{\rm ph} = 0$.
Staying within the homogeneous regime, and setting the left hand side of Eq.~\eqref{eq:KineticEq} to zero while using Eq.~\eqref{eq:I_Lead} for $I^{\rm tun}_{k\alpha}$, we obtain
\begin{equation}
\tilde{F}_{k\alpha} = \frac{\sum_n \Gamma^n_{k\alpha} D(\mc{E}_{k\alpha}^n)}{\sum_n \Gamma^n_{k\alpha}}.
\label{eq:reservoirSteadyState}
\end{equation}
For a typical metallic reservoir with a wide bandwidth (greater than $\hbar\Omega$), the photon-assisted tunneling rates $\Gamma^n_{k\alpha}$ in Eq.~\eqref{eq:FFGR_Lead} may be significant for $n \neq 0$.
Consequently, the sum over $n$ in Eq.~\eqref{eq:reservoirSteadyState} leads to steady state occupations which differ substantially from those given by a simple Floquet-Fermi-Dirac distribution $F_{k\alpha}=D(\mathcal{E}_{k\alpha})$.

\begin{figure}[t]
\begin{center}
\includegraphics[width=0.95\columnwidth]{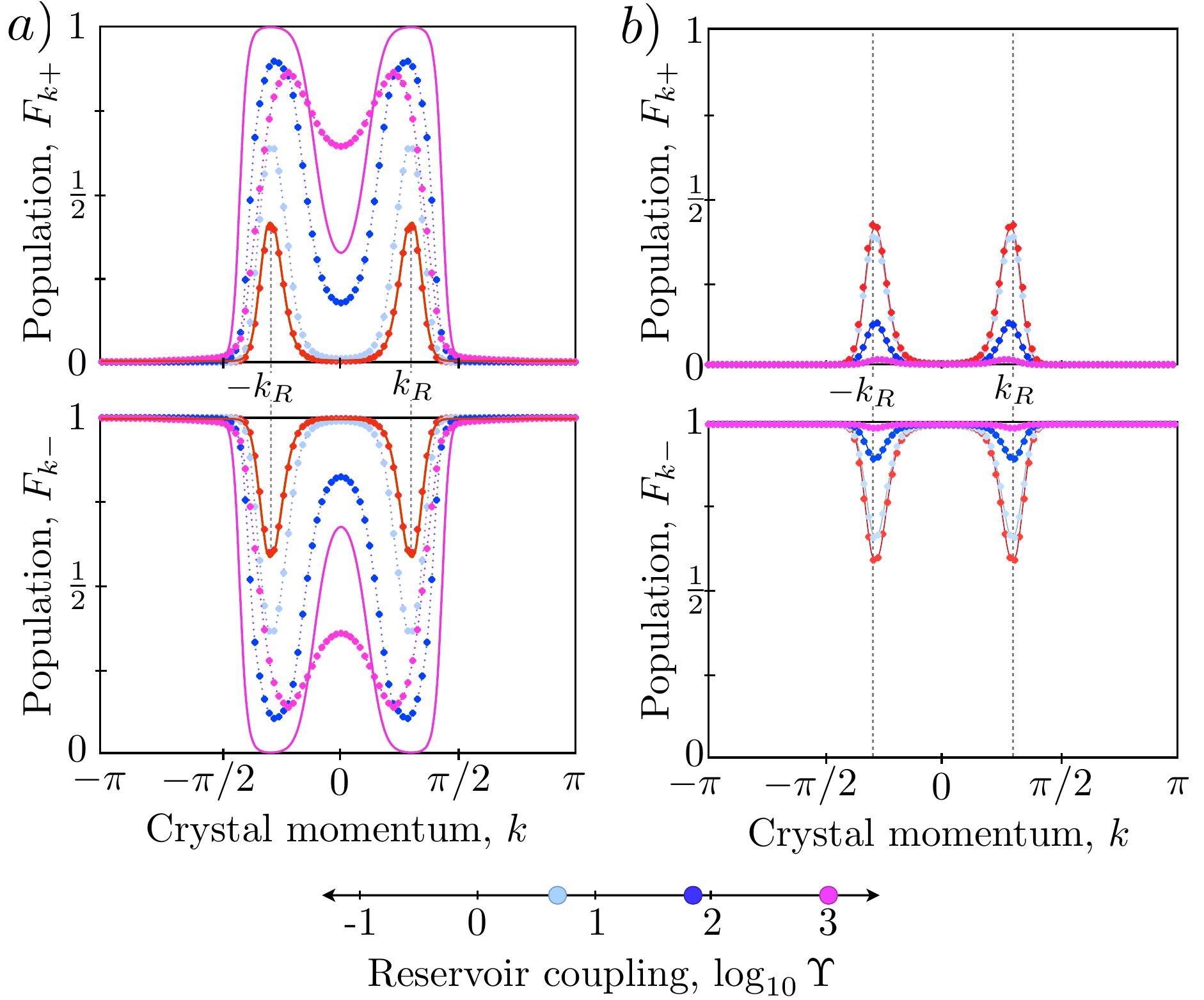}
\caption{Numerically obtained steady states of the system coupled to both bosonic and fermionic baths.
The top and bottom panels show the distributions of electrons in the Floquet + and - bands, respectively, for increasing strength of the coupling to the Fermi reservoir.
We characterize the coupling strength by the ratio of tunneling and recombination rates, {$\Upsilon = 2\Gamma^0_{k_R,+}/\mathcal{W}^{\rm rec}_{k=0}$ }(see Eqs.~(\ref{eq:gamma_rec}) and (\ref{eq:FFGR_Lead}) for definitions of the rates).
Two types of Fermi reservoirs are studied. (a) \textit{Wide-band} Fermi reservoir, whose Fermi level lies in the middle of the original bandgap (the bandgap of $H_0$).  An increase in the coupling strength to such a reservoir leads to a substantial increase in the electron and hole densities $n_e$ and $n_h$, due to photon assisted tunneling. (b) \textit{Energy filtered} Fermi reservoir,  whose Fermi level lies at the resonance energy $E_0$ in the original conduction band, i.e., in the middle of the Floquet gap of the driven system. The electron and holes densities $n_e$ and $n_h$ are suppressed via the coupling to the narrow-band Fermi reservoir. In all panels, the red data points are for a half filled system which is disconnected from the Fermi reservoir. The other colors correspond to the values of $\Upsilon$ indicated at the bottom.  The solid lines are fits to Floquet-Fermi-Dirac distributions, with  separate chemical potentials for electrons and holes in the Floquet $+$ and $-$ bands, respectively. The temperature taken for the fits is identical to the phonon and reservoir temperature, $k_BT=10^{-2}\hbar\Omega$.  In these simulations, the parameters for the photon (recombination) and phonon baths were kept fixed at the values yielding the green curve in Fig.~\ref{fig:Results1}, while we vary the overall scale of the coupling strength to a homogeneously coupled fermionic reservoir.}
\label{fig:Results2}
\end{center}
\end{figure}

We now directly illustrate the difficulties which arise from coupling the periodically-driven system to a wide-band fermionic reservoir, in the more general case where the system is also subject to electron-phonon coupling and radiative recombination, $I^{\rm rec}$, $I^{\rm ph} \neq 0$.
In Fig.~\ref{fig:Results2}a we plot steady state distributions for several values of the coupling strength to the reservoir.
The parameters of the bosonic bath (phonons and recombination) are held fixed, with values identical to those yielding the green (middle) curve of Fig.~\ref{fig:Results1}a.
We start at half filling, with the chemical potential of the reservoir set in the middle of the bandgap of the non-driven system, i.e.,~we set $\mu_{\rm res} = E_0 - \hbar\Omega/2$, see Fig.~\ref{fig:setup}. The system-reservoir coupling $J_{\ell,k\nu}$ as well as the reservoir density of states are taken to be constant \cite{footnote:widebandwidth}. As the reservoir coupling increases, the steady state distribution becomes ``hotter'', with a higher and higher density of excitations.

The heating effects of the reservoir can be understood as follows.
In terms of the original (non-driven) bands, Fig.~\ref{fig:setup}b, the leading order effect of the reservoir is to populate valence-band-like states and to empty conduction-band-like states.
In terms of the Floquet bands, this in particular entails  removing electrons from states in the lower Floquet band within the momentum window $-k_R < k < k_R$, and injecting electrons into states of the upper Floquet band in the same momentum window.
Qualitatively, this is similar to the effect of radiative recombination, compare to Fig.~\ref{fig:Results1}.
Strong coupling to the reservoir thus leads to a large density of excitations in the Floquet bands.
To achieve an insulator-like distribution, as needed for  the realization of a Floquet topological insulator, these excitations must be suppressed.

\subsection{Energy filtered fermionic reservoirs}
\label{subsec:filtering}

Interestingly, photon-assisted tunneling can be effectively suppressed if the system-reservoir coupling is mediated through a narrow band of ``filter'' states (realizations are discussed below).
For illustration, let us imagine the system connected via tunneling to an energy filter: a device with a finite density of states in a restricted energy range, whose states couple strongly to the electron reservoir. The filter states hybridize with the reservoir states to produce a peak in the continuum density of states within the filter energy window.
If the system is only coupled to the fermionic reservoir via the filter states, then the effective tunneling density of states (TDOS) that enters in the transition rates via Eq.~\eqref{eq:FFGR_Lead} quickly falls to zero outside the filter window (Fig.~\ref{fig:filter}c). Note that in the above discussion we assumed that the filter is not subject to the external drive\cite{footnote:filterDriving}.

As a concrete example, consider resonant tunneling through a single filter level at energy $E_{\rm filter}$.
Here we find tunneling rates with a Lorentzian dependence on energy: $\Gamma^n \sim \gamma/[(\mathcal{E}_{k\alpha} + n\hbar\Omega - E_{\rm filter})^2 + (\hbar\gamma/2)^2]$, where $\gamma$ is the level broadening of the filter state due to its coupling to the continuum of reservoir modes. Consider, for example, setting $E_{\rm filter}=\frac12\hbar\Omega=E_0$ (i.e.,~at the conduction band resonance energy). Then, in the limit $\Omega \gg \gamma$, photon-assisted tunneling rates (${n\neq 0}$) are strongly suppressed.
If the energy filter consists of multiple resonant levels connected in series, or a narrow band of states, a sharper ``box-like'' transmission window can be obtained (see, e.g.,~Ref.~\onlinecite{Whitney2014}).

In practice, the energy filter may be realized by coupling the system to the reservoir via a section of large bandgap material hosting a narrow band of impurity states deep inside its gap.
The intermediate band should satisfy three essential requirements: (i) The Fermi level should lie inside it, (ii) The band should be narrower than the width $\hbar \Omega$ of a single Floquet zone, as discussed above, and (iii) it should be separated from the conduction and valence bands of the host material by more than $\hbar \Omega$, to avoid the direct absorption of photons from the driving field.
Highly mismatched alloys featuring narrow bands of extended states in their bandgaps have been realized in the context of intermediate-band solar cells~\cite{Yu2003,Lopez2011,Luque2012}. 
 We expect similar methods to allow for the realization of the energy filter introduced in this work.
Energy filtering through quantum dots could provide an alternative approach.
Due to their large size as compared to atoms, however, achieving a level spacing exceeding $\hbar \Omega$ may prove challenging (especially at optical frequencies)~\cite{Luque2012}.

\subsection{Steady state with filtered reservoir}
\label{subsec:steadyStateFiltered}

We now investigate how coupling to an energy-filtered reservoir affects the steady state of the system. We start with the case where phonons and radiative recombination are absent, $I^{\rm rec} = I^{\rm ph} = 0$. Throughout the discussion below we assume a box-like filter, such that the tunneling density of states is strictly zero outside the filter window.

When the filter window (bandwidth) is narrower than $\hbar \Omega$, photon-assisted processes are suppressed.
According to Eq.~(\ref{eq:reservoirSteadyState}), the occupation distribution in the reservoir, taken as a function of energy, is then directly mapped into the occupation distribution of the driven system, taken as a function of {\it quasi-energy} (i.e.,~the occupation $F_{k\alpha}$ of each Floquet state $\Ket{\psi_{k\alpha}}$ is determined by a single term, $D(\mathcal{E}_{k\alpha})$ with fixed $n$, on the right hand side).
In the case of the wide-band reservoir, half-filling was ensured by placing the chemical potential of the reservoir  in the middle of the gap of the non-driven system.
Here, the best choice is to center the filter window around the energy of the resonance in either the conduction band or the valence band, $\mu_{\rm res} = \pm\hbar\Omega/2$, and also to set the chemical potential $\mu_{\rm res}$ close to the resonance value (for the simulations below, we center the filter window around the resonance in the conduction band).
In this way, the reservoir chemical potential will end up inside the Floquet gap. Note that although the chemical potential $\mu_{\rm res}$ is set to an energy within the conduction the bands of the non-driven system, the filtering prevents a large inflow or outflow of electrons which would otherwise push the density far away from half-filling. Due to the fact that this is a highly non-equilibrium situation, however, some density shifts away from half-filling are generically expected (see discussion below).

The ideal Floquet insulator distribution can be achieved in the situation where the reservoir chemical potential is set inside the Floquet gap, and where the filter window is wide enough to cover the full bandwidth of the Floquet-Bloch band structure, but narrower than the driving field photon energy $\hbar\Omega$ such that photon-assisted processes are still suppressed.
More generically, however, the filter window will be narrower than the bandwidth of the Floquet-Bloch bands, as depicted in Fig.~\ref{fig:setup}c.
In this case the kinetic equation~\eqref{eq:KineticEq} with $I^{\rm rec} = I^{\rm ph} = 0$ does not have a unique steady state, as excited electrons and holes above and below the filter edges, respectively, have no way to relax.
However, this is an unstable situation: any small scattering rate due to acoustic phonons will allow electrons to ``trickle down'' and fill up all Floquet states below the bottom of the filter window. In the absence of Floquet-Umklapp processes, the resulting steady state will correspond to that of an insulator at finite temperature (assuming the same temperature for the phonons and the Fermi reservoir).
More specifically, the electronic distribution for both bands will be described by a global Floquet-Fermi-Dirac distribution with a single chemical potential, which is set by that of the reservoir.

Once we include the contributions of Floquet-Umklapp processes such as recombination, the steady state hosts densities of excited electrons and holes, $n_e$ and $n_h$, respectively (which are generally large compared with the thermal density $n_{\rm th}$).
In the limit of a weakly coupled reservoir, the combined density of electrons and holes $\bar{n} = n_e + n_h$ is determined solely by the recombination and phonon scattering rates, as discussed in Sec.~\ref{sec:bosons}.
The steady state excitation density is further suppressed with increasing coupling to the reservoir, as we demonstrate below (see Fig.~\ref{fig:Results2}b).

\begin{figure}[t]
\begin{center}
\includegraphics[width=1.0\columnwidth]{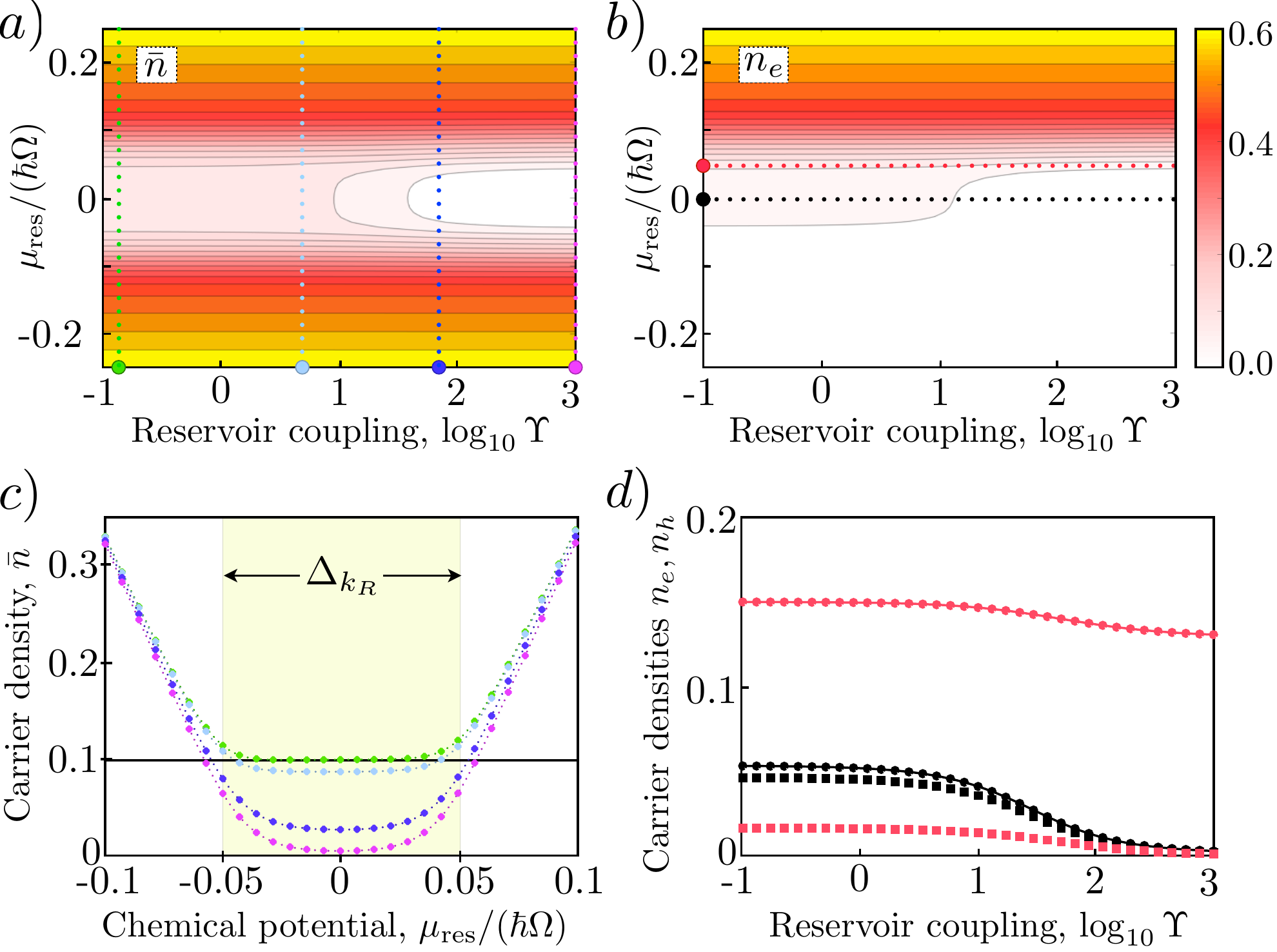}
\caption{Electron and holes densities $n_e$ and $n_h$ in the steady state of the system coupled to bosonic baths (acoustic phonons and recombination) and an \textit{energy filtered} fermionic reservoir. The figure clearly demonstrates that (1) the steady state densities $n_e$ and $n_h$ are insensitive to small shifts of the reservoir's chemical potential $\mu_{\textrm{res}}$ near the middle of the Floquet gap; and (2)  a sufficiently strong coupling to the reservoir can effectively suppress the electron and hole densities when $\mu_{\textrm{res}}$ is within the Floquet gap.
Panel (a) shows the total density $\bar{n}=n_e+n_h$ as a function of the Fermi level of the reservoir $\mu_\textrm{res}$ and the coupling strength ratio {$\Upsilon = 2\Gamma^0_{k_R,+}/\mathcal{W}^{\rm rec}_{k=0}$}.  As long as $\mu_\textrm{res}$ is within the Floquet gap, $n_e$ and $n_h$ remain low. Once $\mu_\textrm{res}$ enters the Floquet $+$ or $-$ bands, the system becomes metallic and the electron (hole) density $n_e$ ($n_h$) is set by the Fermi level of the reservoir. 
This behavior is seen in panel (b), where we plot $n_e$. To further demonstrate the incompressible regime, in (c) we show $\bar{n}$ as a function of $\mu_\textrm{res}$  for several coupling strengths to the reservoir, corresponding to the dotted lines in panel (a).  Panel (d) gives the the electron and hole densities, $n_e$ (circles) and $n_h$ (squares) for two values of $\mu_\textrm{res}$: in the middle of the Floquet gap (black) and at the edge of the $+$ Floquet band (red). In the first case, the results explicitly demonstrate the suppression of the excitation densities $n_e$ and $n_h$  with increasing reservoir coupling. Model parameters are the same as in Fig.~\ref{fig:Results2}.}
\label{fig:Results3}
\end{center}
\end{figure}

While the steady state electron and hole excitation densities are equal for a half-filled system without coupling to a Fermi reservoir (Sec.~\ref{sec:bosons}),
 $n_e$ and $n_h$ need not be equal when  the reservoir is present, even when the chemical potential of the filtered reservoir is placed in the middle of the Floquet gap.
To see why, note that here the Fermi level of the filtered reservoir is aligned with the resonance energy $\frac12\hbar\Omega$ in the conduction band of the non-driven system.
The asymmetric placement of the energy window of the reservoir with respect to the non-driven band structure generically breaks any effective particle-hole symmetry, and yields a shift of the total density away from half-filling.
Importantly, the shift $\Delta n = n_e-n_h$ can be small, being bounded by $\bar{n}$.
More careful considerations (see Appendix~\ref{appendix: asymmetry}) show that $\Delta n$ is in fact expected to be significantly lower than $\bar{n}$, which is confirmed by our numerical simulations (see Fig.~\ref{fig:Results3}d).

Staying within the regime of a weakly coupled reservoir, let us now consider what happens when the reservoir's Fermi energy is shifted away from the center of the Floquet gap. As long as the Fermi level of the reservoir remains within the Floquet gap, the occupation factors $D(\mathcal{E}_{k\alpha})$ in Eq.~(\ref{eq:I_Lead}) change only weakly, due to the finite temperature of the reservoir. Since the rates $\Gamma^n_{k\alpha}$ are independent of the occupation of the reservoir, the changes in $I^{\textrm{tun}}_{k\alpha}$ are only ``thermally activated'' by the reservoir's temperature. We therefore expect the steady state of the system to be only weakly affected.  This implies that an interesting situation has been obtained, in which the driven system becomes \textit{incompressible}, with respect to changes of the reservoir's Fermi level. Once the Fermi level enters, say, the upper Floquet band, the density of excited electrons in the band is greatly affected. If we approximate the distribution of excited electrons by a Fermi function, we can expect its Fermi level to track the Fermi level of the reservoir.

As the strength of the coupling to the reservoir is increased, we expect the Fermi reservoir to become more dominant in setting the steady state of the system. In the limit where the coupling to the reservoir dominates all other scattering mechanisms, we expect the steady state to be described by a global Floquet-Fermi-Dirac distribution, with the same chemical potential as that of the filtered reservoir. Note that in this limit, a non-zero coupling to the phonon bath is still important in order to allow electrons to fill up states from the bottom of the lower Floquet band up to the reservoir's Fermi level.

The above considerations are confirmed by our numerical simulations, which are given in Figs.~\ref{fig:Results2} and~\ref{fig:Results3}.  In these simulations, we fix the matrix elements describing the coupling to the photon (recombination) and acoustic phonon baths as in the green (middle) curve of Fig.~\ref{fig:Results1}a, and vary the overall scale of the couplings $J_{\ell,k\nu}$ to the Fermi reservoir (which are taken to be uniform). The reservoir density of states is taken to be constant in a window of width $\hbar\Omega/2$ placed symmetrically around $E_0 = \frac12 \hbar\Omega$. The distributions of electrons $\langle f^\dagger_{k\alpha}f_{k\alpha}\rangle$ in the two bands are plotted in Fig.~\ref{fig:Results2}, and are separately fitted to Floquet-Fermi-Dirac distributions with independent chemical potentials $\mu_e$ and $\mu_h$ for electrons and holes, as in Sec.~\ref{subsec: bosons steady}.

Figure~\ref{fig:Results2} clearly demonstrates that for a wide-band reservoir, panel (a), the density of excitations \textit{increases} when the coupling to the reservoir is increased; in contrast, for a filtered reservoir,  panel (b), the density of excitations \textit{decreases} with increasing coupling strength to the reservoir. For the filtered reservoir, the chemical potential sits at the resonance energy in the conduction band, $\frac12 \hbar\Omega$. In all fits in Fig.~\ref{fig:Results2} we set the temperatures of the  Floquet-Fermi-Dirac distributions to be identical to the phonon and reservoir temperature. While we obtain excellent fits at weak reservoir coupling, the fits become less accurate when the coupling to the reservoir is increased.
This arises due to the non-uniform way in which the reservoir is coupled to the Floquet bands.
 As in Sec.~\ref{subsec: bosons steady}, we verify that the scattering rates in the numerically obtained steady states are significantly smaller than the Floquet gap $\Delta_{k_R}$ for all reservoir coupling strengths used (see Appendix \ref{appendix: rates}).

In Fig.~\ref{fig:Results3} we study the densities $n_e$ and $n_h$ as functions of the strength of the coupling to the reservoir and its chemical potential.
The figure  demonstrates two important points.
First, the steady state densities $n_e$ and $n_h$ are insensitive to small shifts of the Fermi level of the reservoir away from the middle of the Floquet gap, yielding an ``incompressible'' behavior $dn_{e,h}/d\mu_{\rm res} \approx 0$. This is demonstrated most clearly by panel (c), which shows $\bar{n}$ vs. $\mu_{\textrm{res}}$ (similar plots of $\Delta n$ can be found in  Appendix~\ref{appendix: asymmetry}).
Second, when the Fermi level of the reservoir lies within the Floquet gap, a sufficiently strong coupling to the reservoir can effectively suppress the electron and hole densities, giving a steady state close to an ideal Floquet insulator.

The coupling strength at which the reservoir is expected to significantly affect the steady state excitation density can be estimated as follows.
Radiative recombination acts to increase the excitation density with the rate $\dot{n}_e^{\rm rec} = \gamma^{\rm rec}$ defined in Eq.~(\ref{eq:gamma_rec}).
As above, we approximate $\gamma^{\rm rec}$ by $\gamma^{\rm rec} \approx (k_R/\pi) \overline{\mathcal{W}}^{\rm rec}$, where $k_R/\pi$ represents the fraction of states that participate in the recombination process and $\overline{\mathcal{W}}^{\rm rec}$ is the average recombination rate in this interval.
Meanwhile, the reservoir can extract excitations at a rate $\dot{n}^{\rm tun}_e \approx - \Gamma^0 n_e$, where $\Gamma^0 \equiv \Gamma^0_{k_R,+}$ characterizes the rate for a single electron to tunnel in or out of the system.
When the reservoir is weakly coupled, the steady state excitation density is controlled by the rate of phonon-mediated interband relaxation, as discussed around Eq.~(\ref{eq:gamma_rec}).
The reservoir begins to play an important role when $\Gamma^0 n_e \gtrsim (k_R/\pi) \overline{\mathcal{W}}^{\rm rec}$, or equivalently when
{\begin{equation}
\Upsilon \equiv (2\Gamma^0/\mathcal{W}^{\rm rec}_{k=0}) \gtrsim k_R/(\pi n_e).
\end{equation}

This relation is indeed borne out in Fig.~\ref{fig:Results3}d, where $n_e \approx 0.05$ and $k_R/\pi \approx 0.3$, and the excitation density is suppressed for $\log_{10} \Upsilon \gtrsim 1$.
Note that Fig.~\ref{fig:Results3}d
also shows a small non-zero value of $\Delta n \ll \bar{n}$ when the Fermi level is in the middle of the Floquet gap (black symbols in Fig.~\ref{fig:Results3}d), arising from the asymmetry of the coupling of the reservoir to the two Floquet bands.

\subsection{Point coupling to a lead}

In many experimentally relevant situations, the system is coupled to a lead at a single point. What is the spatial dependence of the steady state in this situation?
So far we have discussed {\it homogeneous} steady state distribution functions $\{\tilde{F}_{k\alpha}\}$, which depend on momentum and band indices but not on position. A {\it homogeneous} steady state can arise in several situations. For the bosonic baths discussed earlier, we assumed a uniform coupling throughout the system. Therefore by themselves the bosonic baths yield a spatially homogeneous steady state distribution. Clearly, if in addition we introduce a fermionic reservoir which is coupled homogeneously throughout the system, a spatially homogeneous steady state is expected.
In addition, for a lead coupled at a single point, there are still two limits in which the steady state remains uniform: (1) absence of any other sources of dissipation; and (2) the limit of a small system size. In the latter case, a uniform distribution is obtained when the level spacing of the system's single particle states is larger than the tunneling rate to and from the reservoir; this corresponds to a tunneling time that exceeds the time required for an electron to traverse the system.

For larger system sizes, where the stringent criterion above is not met, the steady state need not 
be spatially homogeneous. If the tunneling rates are comparable to or larger than the level spacing, the coupling to the reservoir can yield nonzero values of the ``coherence'' terms $\langle f^\dagger_{k\alpha}f_{k'\beta}\rangle$, which generically cause spatial inhomogeneity.
Calculating the full set of such coherences is a formidable task.
Fortunately,  we can gain an intuitive understanding of the form of the inhomogeneous steady state by considering the dynamics of a spatially-dependent excitation densities $n_e(x, t)$ and $n_h(x,t)$.
Close to the lead, placed at $x = 0$, the excitation density will be affected by the lead and will roughly correspond to that found for a homogeneous system-reservoir coupling.
Far from the lead, we expect the excitation densities to relax to bulk values $n^{\rm bulk}_e$ and $n^{\rm bulk}_h$.
Below we estimate the ``healing length'' over which this transition occurs.

Due to fast intraband electron-phonon scattering (which is still slow compared with the driving frequency and the on-resonance Rabi frequency $\Delta_{k_R}/\hbar$), carrier motion on time scales much larger than the driving period is expected to be diffusive.
The corresponding diffusion constant can be estimated as $D = \bar{v}^2/W^{\rm intra}$, where $\bar{v}$ is a typical velocity of the excitations~\cite{footnote:observables} and $W^{\rm intra}$ is a typical \textit{intraband} scattering rate from acoustic phonons, both taken in the steady state.
Focusing on the situation near half-filling and incorporating the source and sink terms due to recombination and interband phonon scattering discussed in Sec.~\ref{sec:bosons}, we obtain two reaction-diffusion equations for the electron and hole densities, $\partial_t n_\lambda= D\,\partial^2_x n_\lambda + \gamma^{\rm rec}\, - \,\Lambda^{\rm inter}n_en_h$, with $\lambda=e,h$.
Adding and subtracting these equations, we find the reaction-diffusion equations
governing the total and offset densities $\bar{n}$ and $\Delta n$,
\begin{eqnarray}
\label{eq:Diffusion}
\partial_t \bar{n} &=& D\,\partial^2_x \bar{n}+ 2\gamma^{\rm rec} - \tfrac{1}{2}\Lambda^{\rm inter}\left(\bar{n}^2-\Delta n^2\right),\nonumber\\
\partial_t \Delta n&=&  D\,\partial^2_x \Delta n.
\end{eqnarray}

For the boundary conditions for the above equations, we use $\Delta n(x=0)$ and $\bar{n}(x=0)$ which are set by the lead, as well as $\partial_x \Delta n = \partial_x \bar{n}=0$ for $x\gg0$, which corresponds to no net flow of particles into the system. Eq.~\eqref{eq:Diffusion} entails two main consequences for the spatial distribution of the steady state, $\partial_t n_\alpha(x,t) = 0$. First, the shift of the total density of electrons from half filling, $\Delta n$, is in fact homogeneous across the system, and set by the lead. Furthermore, linearizing Eq.~\eqref{eq:Diffusion} around the bulk steady state gives the healing length
\be
\xi = \left(\frac{D \bar{n}^{\rm bulk}}{4\gamma^{\rm rec}}\right)^{\!1/2},
\label{eq:healing}
\ee
where we can approximate $\bar{n}_{\rm bulk} = 2\sqrt{k_R \overline{\mathcal{W}}^{\textrm{rec}}/\pi \Lambda^{\rm inter}}$ from Eq.~\eqref{eq:rho_steady}.
Here we neglect corrections due to a small $\Delta n$.
For system sizes smaller than $\xi$, a lead coupled at a point can be effective in setting the distribution throughout the system.
In such a system, for a sufficiently strong point coupling to a filtered lead, a Floquet insulator distribution can be achieved, as was shown for the homogeneous case in Sec.~\ref{subsec:steadyStateFiltered}.

\section{Summary and Discussion}
\label{sec:discussion}

The ability to control and probe non equilibrium quantum many body systems poses one of the most outstanding challenges in modern condensed matter physics.
In this paper, we analyzed steady states in a model for a periodically driven semiconductor, and demonstrated the means through which these steady states can be controlled.  
We considered the open system dynamics of a resonantly driven electronic system coupled to acoustic phonons and the electromagnetic environment, as well as an external fermionic reservoir. 
The couplings to these baths have two complementary roles: they allow energy relaxation, but may also induce processes which lead to heating.
Motivated by the prospect of realizing Floquet topological insulators, our goal was to find the conditions under which the steady state resembles a band insulator.
Importantly, we focused on the regime where the scattering rates in the steady state are smaller than the drive-induced Floquet gap. 
Only in this regime could we expect to observe effects requiring quantum coherence, such as drive-induced topological phenomena. 

Starting with the case where the system is coupled only to the bosonic baths, we have shown that the system can approach a Floquet insulator steady state with an added density of excitations in the two Floquet bands.
The density of excitations is controlled by the ratio of radiative recombination and electron-phonon scattering rates, and can be small for experimentally-relevant parameter values.
We found a square root dependence of the excitation density on the above ratio, see Eq.~(\ref{eq:rho_steady}), which implies that additional controls are needed to fully suppress the deviations from the Floquet insulator state.
Next, we considered the effects of coupling to an external Fermi reservoir, which plays two important roles in our setup.
First, the reservoir is a crucial component for transport experiments.
Importantly, we show this coupling significantly increases the density of excitations, unless the reservoir is coupled through an energy filter. 
Second, the energy-filtered reservoir can serve as an additional control to {\it reduce} the density of excitations, bringing the system closer to the ideal Floquet insulator state.

A main motivation for our work is the prospect of obtaining a Floquet topological insulator: a driven system with an insulating bulk but with conductive edge and surface modes. 
What are the implications of our results for transport?
Interestingly, our results show that even in the presence of a finite density of excitations, the steady state of the driven system can exhibit an ``incompressible'' behavior: the steady state is unaffected by small changes in the chemical potential of the {\it energy filtered} reservoir, as long as it situated near the middle of the Floquet gap~\cite{footnote:finiteTemp}. 
This behavior is shown in Fig.~\ref{fig:Results3}.
In addition, when energy filtered leads are used for transport, photon-assisted conduction channels are suppressed.
In the case of neutral particles, the incompressibility implies insulating behavior: no current would result from a small source-drain bias between two spatially-separated energy-filtered leads. 
This follows from the insensitivity of the steady state to the leads' chemical potentials.

When we consider charged particles, however, the non-zero density of excited carriers is expected to yield a finite resistivity even in the paramter range corresponding to the insulating regime above. Given the diffusive nature of particle motion in the system, we estimate the local resistivity $\sigma^{-1}(x)$ using the Drude form: $\sigma^{-1}(x) = |m_*|W^{\rm intra}/[e^2 \bar{n}(x)]$, where $m_*$ is the effective mass around the Floquet upper (lower) band minimum (maximum), $e$ is the electric charge. 
Consider now a two-terminal transport measurement using energy filtered leads, through such a system of charged carriers. If the system is small enough such that the steady state is spatially homogeneous, a sufficiently strong coupling to energy filtered leads can, in fact, suppress the density of excited carriers and yield nearly-insulating behavior. For larger systems, with a spatially inhomogeneous steady state, the total resistance $R$ is given by the sum of series resistances, $R = \int_0^L \,dx  \sigma^{-1}(x)$.
The bulk of the system gives an extensive contribution $R_{\rm bulk}\approx L |m_*|W^{\rm intra}/\left[e^2 \bar{n}^{\textrm{bulk}}\right]$.
Interestingly, if the lead coupling is strong, the excitation density near the ends of the system will become very small and thus give a large contribution $R_{\rm end}$ to the resistance. Therefore, for a fixed system size, the system may obtain insulating behavior in a two-terminal measurement upon increasing the coupling strength of the lead.

The analysis of steady states in driven electronic systems is currently the subject of intense activity (see e.g., Refs.~\onlinecite{Dehghani2014, Iadecola2014}). 
In Ref.~\onlinecite{Dehghani2014}, time evolution and steady states after a quench were studied for a 2D semiconductor with a topological Floquet spectrum.
There, the authors considered open system dynamics with momentum conserving interactions with a bosonic bath, and found regimes exhibiting quantized Hall conductivity.
In addition, Ref.~\onlinecite{Iadecola2014} studied a resonantly-driven electronic system where the only relaxation pathway was through an external fermionic lead, and found that a grand canonical distribution could be obtained under finely-tuned conditions. 
In our work, which was carried out in parallel, we considered the combined effects of momentum and energy relaxation through the coupling to acoustic phonons and the coupling to an external fermionic reservoir.
Notably, we included the inevitable effects of heating due to Floquet-Umklapp processes, as exemplified by radiative recombination.
Importantly,  momentum relaxation played a crucial role in establishing the Floquet insulator steady state under these conditions.

To make a connection with experimentally relevant regimes, we compare our model parameters with those accessible in solid state systems. 
Consider a drive frequency of $\Omega=2\pi\times 100$ THz, which translates to $0.4$ eV in energy units. Correspondingly, the parameters used in Sec.~\ref{sec:bosons},
yield a Rabi frequency (Floquet gap) of $\Delta_{k_R}/\hbar\approx 2\pi \times 10$ THz (translating to $40$ meV), and
a characteristic phonon relaxation time scale of $[\sum_{k'}W_{k=0,+}^{k'+}]^{-1}=1 \;\textrm{ps}$.
This is the total relaxation rate out of the state $k = 0$ in the upper band. The recombination
time scales, $(\mathcal{W}_{k=0}^{\textrm{rec}})^{-1}$ used to obtain the steady state distribution in Fig.~\ref{fig:Results1}, are {[}$1$ $\mu$s, $60$ ns, $3$ ns, $180$ ps, $10$ ps{]}. 
When coupling to the fermionic reservoir was introduced in Sec.~\ref{sec:lead}, we fixed  the recombination time scale at $3$ ns; the steady state distributions
in Fig.~\ref{fig:Results2} correspond to tunneling times $(\Gamma_{k=0,-})^{-1}$ of approximately {[}$200$ ps, $30$ ps, $3$
ps{]}. 
In Fig.~\ref{fig:Results3}, the tunneling time from the reservoir varies
from $30$ ns to $3$ ps. 
Note that these values are in line with those in typical semiconductor nanostructures, where tunneling times can vary widely\cite{Shah99}.

Although our model is inspired by resonantly-driven semiconductors, we expect our conclusions and formalism to be relevant to a broad variety of driven-dissipative systems including cold atomic gases. Our results also have important implications for Floquet topological insulators. Indeed, they provide a roadmap towards the practical realization
of the Floquet insulator state, which is key to observing quantized transport in Floquet topological insulators. 
We expect engineered reservoirs of carriers to be particularly useful in this context, allowing to perform transport measurements while stabilizing insulating-like steady states.

Several aspects of the problem require further study. In this work we have not addressed the effect of inter particle interactions.  Floquet-Umklapp processes involving inter-particle scattering give an additional channel for the system to absorb energy from the driving field and increase the number of excitations.  However, our current work demonstrates that coupling to a bath of phonons, can keep the heating and excitation density under control. Another important direction is a careful study of the inhomogeneous steady states of Floquet topological insulators, which is crucial in order to predict the edge and surface response of these systems.

The authors would like to thank A. \.Imamo\u glu, C. Grenier, A. Srivastava, and L.I. Glazman for insightful discussions. Financial support from the Swiss National Science Foundation (SNSF) is gratefully acknowledged. MR acknowledges support from the Villum Foundation and from the People Programme (Marie Curie Actions) of the European Union's Seventh Framework Programme (FP7/2007-2013) under REA grant agreement PIIF-GA-2013-627838. NL acknowledges support from the Israel-US Binational Science Foundation, and I-Core, the Israeli excellence center ``Circle of Light''. GR and KS are grateful for support from NSF through DMR-1410435, as well as the Institute of Quantum Information and matter, an NSF Frontier center funded by the Gordon and Betty Moore Foundation, and the Packard Foundation.

\appendix

\section{Kinetic equation in the Floquet basis}
\label{appendix: kinetic}

In this appendix we discuss the key points in the derivation of the Floquet kinetic equation, represented schematically in Eq.~(\ref{eq:KineticEq}) of the main text.
Specifically, the aim is to derive a system of differential equations which describe the time evolution of the Floquet state occupation factors $F_{k\alpha}(t) = \avg{f^\dagger_{k\alpha}(t)f_{k\alpha}(t)}$, where $f^\dagger_{k\alpha}(t)$ and $f_{k\alpha}(t)$ are the creation and annihilation operators for Floquet states defined above Eq.~(\ref{eq:KineticEq}).
The time derivative $\dot{F}_{k\alpha} = \frac{d}{dt}\avg{f^\dagger_{k\alpha}(t)f_{k\alpha}(t)}$ couples to an infinite hierarchy of higher and higher order correlation functions.
The main approximation is to truncate this hierarchy at the lowest non-trivial order, and obtain a closed system of evolution equations.
We now outline the required steps.

\subsection{Basis transformation and dressed matrix elements}

The first important step is to express the electronic terms in the Hamiltonian in terms of the Floquet creation and annihilation operators.
Using Eq.~(\ref{eq:FloquetState}), the transformation is made via
\begin{eqnarray}
\nonumber c_{k\nu}^{\dag} & = &
\sum_{\a}\sum_{n}e^{i(\s E_{k\alpha}+n\Omega)t}\qb{\phi_{k\alpha}^{n}}{\nu k}f_{k\a}^{\dag}(t),\\
c_{k\nu} & = &
\sum_{\beta}\sum_{m}e^{-i(\s E_{k\beta}+m\Omega)t}\qb{\nu k}{\phi_{k\beta}^{m}}f_{k \beta}(t),
\end{eqnarray}
where $\kett{\nu k}$ is the Bloch function corresponding to crystal momentum $k$ in band $\nu$ of the non-driven system, and $f_{k\alpha}^{\dag}(t)$ and $f_{k\a}(t)$ are creation and annihilation operators for the Floquet state $\kett{\psi_{k\a}(t)}$ (we simplify notations by setting $\hbar=1$ here and everywhere below).
Using these relations, we write the electron-boson interaction and system-reservoir tunneling Hamiltonians as
\begin{widetext}
\begin{eqnarray}
\label{eq:HintFloquetBasis} H_{\rm int} & = & \sum_{k\vec{q}}\sum_{\alpha\alpha'}\sum_{n}e^{i(\s E_{k-q_x,\alpha'}-\s E_{k\alpha})t}e^{in\Omega t}\mathcal{G}^{(n)}_{\alpha'\alpha}(k, q_x) f_{k-q_x,\alpha'}^{\dag}(t)f_{k\alpha}(t)(b_{-\vec{q}}+b_{\vec{q}}^{\dag}),\\
\label{eq:HtunFloquetBasis} H_{\rm tun} & = & \sum_{k\ell}\sum_{n}\left(e^{i\s E_{k\alpha}t}e^{in\Omega t}\mathcal{J}^{(n)}_{\ell, k\alpha}f_{k\alpha}^{\dag}(t)d_{\ell}\ +\ {\rm h.c.}\right),
\end{eqnarray}
with ``dressed'' matrix elements
\begin{eqnarray}
\label{eq:GFloquet} \mathcal{G}^{(n)}_{\alpha'\alpha}(k, q_x) & = & \sum_{\nu}\sum_{m}G^{k-q_x \nu'}_{k \nu}(q_x) \qb{\phi_{k-q_x,\alpha'}^{m + n}}{\nu', k-q_x}\qb{\nu k}{\phi_{k\alpha}^{m}},\\
\label{eq:JFloquet} \mathcal{J}^{(n)}_{\ell, k\alpha} & = & \sum_{\nu}J_{\ell, k\nu}\qb{\phi_{ak}^{n}}{\nu k}.
\end{eqnarray}
\end{widetext}
Here we use $\mathcal{G}$ and $\mathcal{J}$ to indicate the 
coupling matrix elements in the Floquet basis.

Note that in Eq.~(\ref{eq:GFloquet}) we have imposed lattice momentum conservation of the electron-phonon interaction, $G^{k' \nu'}_{k \nu}(q_x) \propto \delta_{q_x,k-k'}$.
Equations (\ref{eq:HintFloquetBasis}) and (\ref{eq:GFloquet}) arise from Eq.~(\ref{eq:ElectronBoson}) of the main text.
Sets of equivalent scattering processes can be identified based on the following useful relation:
\begin{equation}
\mathcal{G}^{(n)}_{\alpha'\alpha}(k, q_x)  = [\mathcal{G}^{(-n)}_{\alpha\alpha'}(k+q_x, -q_x)]^*.
\end{equation}
The explicit form of the dressed matrix elements shows that the overlaps $\braket{\phi_{k\alpha}^m}{k\nu}$ between the original states and the Fourier components $\kett{\phi_{k\alpha}^m}$ of the Floquet modes are crucial in determining the rates of the different Floquet scattering processes, as discussed in the main text.

\subsubsection*{Matrix elements for Floquet-Umklapp processes}

An interesting situation occurs when the coupling to the driving field is defined by a vector $\mathbf{g}$ [see Eq.~(\ref{eq:Hamiltonian})] such that  $\tilde{g}_\parallel=0$ [see Eq.~(\ref{eq: Hamiltonian2}) for the definition of $\mathbf{\tilde{g}}]$. This commonly occurs in experimentally relevant materials driven by optical fields.
Here, the Fourier harmonics $\kett{\phi_{k\alpha}^m}$ have a fixed band character for $m$ of fixed parity: for example, in the convention used throughout the paper and set below Eq.~(3), $\kett{\phi_{k\alpha}^m}$ is proportional to $\kett{k v}$ for $m$ odd and proportional to $\kett{k c}$ for $m$ even, see Fig.~2a.
Additionally, note that scattering by phonons preserves the band character, $G_{\nu\nu'} \propto \delta_{\nu\nu'}$.
As a consequence, under these conditions Floquet-Umklapp processes involving phonons are forbidden for $n$ odd.
Therefore, if in addition to $\tilde{g}_\parallel=0$  the phonon bandwidth is less than the driving frequency $\Omega$, all Floquet-Umklapp processes, including both even and odd $n$, are not allowed.

\subsection{Equations of motion}

We now study the equations of motion for the Floquet state populations $F_{k\alpha} = \avg{f^\dagger_{k\alpha}(t)f_{k\alpha}(t)}$.
The populations are the diagonal part of the ``polarization matrix'' $P_{k\alpha}^{k'\alpha'}(t) = \avg{f^\dagger_{k'\alpha'}(t)f_{k\alpha}(t)}$.
In addition to the populations, this matrix also characterizes coherence between Floquet states with different crystal momenta and/or band indices.
This off-diagonal part  may be important for the dynamics and for characterizing steady states.
In the main text we focus on steady states in a regime where the off-diagonal part of the polarization matrix can be neglected.
Here we derive the kinetic equation in a more general context, including the full polarization matrix, and discuss when and how the off-diagonal parts may be neglected.

As a preliminary, we note the following important property of the Floquet state creation operators $f^\dagger_{k\alpha}(t)$.
Similar relations hold for the annihilation operators.
Let $U(t, t')$ be the single particle time evolution operator corresponding to the Schr\"{o}dinger equation $i\frac{d}{dt}\kett{\psi} = H(t)\kett{\psi}$, with the Hamiltonian $H(t) = \sum_k c^\dagger_{k\nu}H_{\nu\nu'}(k, t) c_{k\nu'}$, where $H(k, t) = H_0(k) + V(t)$ is defined in Eq.~(\ref{eq:Hamiltonian}) in the text.
The operator $f^\dagger_{k\alpha}(t)$ satisfies $f^\dagger_{k\alpha}(t) = U(t, t')f^\dagger_{k\alpha}(t')U^\dagger(t, t')$, which can be written in the differential form:
\begin{equation}
\label{eq:HeisenbergEOM} i\partial_t f^\dagger_{k\alpha}(t) = [H(t), f^\dagger_{k\alpha}(t)].
\end{equation}
This expression will be used below.

The derivation of the kinetic equation proceeds along standard lines, as explained in detail in, e.g.,~Ref.~\onlinecite{Kira2012}.
The main difference from the usual case (i.e., for non-driven systems) is the appearance of the ``dressed'' matrix elements in the interaction Hamiltonians.
Below we set up the calculation and point out where these terms appear, and where special considerations are needed to complete the derivation for the case of a periodically driven system.

We seek the time evolution of the Floquet state populations $F_{k\alpha}(t)$.
However, since these populations are special cases of the polarizations $P_{k\alpha}^{k'\alpha'}(t)$ defined above, for $k = k'$, $\alpha = \alpha'$, we begin with the more general expression for the time derivative of $P_{k\alpha}^{k'\alpha'}(t)$:
\begin{equation}
\label{eq:ddtlevel1}\!\! i\partial_t \avg{f^\dagger_{k'\alpha'}(t)f_{k\alpha}(t)} = \avg{[ f^\dagger_{k'\alpha'}(t)f_{k\alpha}(t), H_{\rm tot} - H(t)]},
\end{equation}
where $H(t)$ is the full single particle Hamiltonian (including driving) as defined above, and $H_{\rm tot} = H(t) + H_{\rm b} + H_{\rm int} + H_{\rm res} + H_{\rm tun}$ is the total Hamiltonian including the baths and the system-bath coupling.
The commutator in Eq.~(\ref{eq:ddtlevel1}) includes two types of contributions, arising  from: (1) the time derivative acting on the state with respect to which the average is taken, and (2) from the explicit time dependence of the operators $f^\dagger_{k'\alpha'}(t)$ and $f_{k\alpha}(t)$.
The latter are given by Eq.~(\ref{eq:HeisenbergEOM}) and its Hermitian conjugate.

To simplify the expressions below, we introduce a more compact notation in which $k$ and the Floquet band index $\alpha$ are compressed into a single index $a$.
In this notation, the dressed electron-phonon coupling matrix elements will be written as $\mathcal{G}^{(n)}_{\alpha'\alpha}(k, q_x) \equiv \mathcal{G}^{(n)}_{a'a}(q_x)$.
The commutator in Eq.~\eqref{eq:ddtlevel1} has two non-trivial terms related to the system-boson and system-reservoir couplings $H_{\rm int}$ and $H_{\rm tun}$, $[f^\dagger_{a}(t)f_{b}(t), H_{\rm int}]$ and $[f^\dagger_{a}(t)f_{b}(t), H_{\rm tun}]$, respectively.
The system-boson coupling produces the following contribution:
\begin{widetext}
\begin{eqnarray}
\ex{[f_{a}^{\dag}(t)f_{b}(t),H_{\rm int}]} & = & \sum_{a'\vec{q}}\sum_{n}e^{i(\s E_{b}-\s E_{a'})t}e^{in\Omega t}\mathcal{G}^{(n)}_{b a'}(q_x)\,\ex{f_{a}^{\dag}(t)f_{a'}(t)(b_{-\vec{q}}+b_{\vec{q}}^{\dag})}\nonumber\\
 & - & \sum_{a'\vec{q}}\sum_{n}e^{i(\s E_{a'}-\s E_{a})t}e^{in\Omega t}\mathcal{G}_{a'a}^{(n)}(q_x)\ex{f_{a'}^{\dag}(t)f_{b}(t)(b_{-\vec{q}}+b_{\vec{q}}^{\dag})},
\label{eq:singletPhononTerm}
\end{eqnarray}
while the system-reservoir coupling leads to
\begin{eqnarray}
 \ex{[f_{a}^{\dag}(t)f_{b}(t),H_{\rm res}]} & = & \sum_{\ell}\sum_{n}e^{i\s E_{b}t}e^{in\Omega t}\mathcal{J}_{\ell,b}^{(n)}\,\ex{f_{a}^{\dag}(t)d_{\ell}}\nonumber\\ &-&\sum_{\ell}\sum_{n}e^{-i\s E_{a}t}e^{-in\Omega t}\mathcal{J}_{\ell,a}^{(n)*}\,\ex{f_{b}(t)d_{\ell}^{\dag}}.
\label{eq:singletReservoirTerm}
\end{eqnarray}
\end{widetext}
Note the appearance of ``mixed'' correlators such as $\ex{f_{a}^{\dag}(t)f_{a'}(t)(b_{-\vec{q}}+b_{\vec{q}}^{\dag})}$ and $\ex{f_{a}^{\dag}(t)d_{\ell}}$ involving both system and bath degrees of freedom, which appear in Eqs.~(\ref{eq:singletPhononTerm}) and (\ref{eq:singletReservoirTerm}).
The expressions are very similar to those that would be obtained for a non-driven system, except that here we find an additional sum over $n$ which accounts for the harmonic structure of the Floquet state wave functions.

In order to describe scattering between Floquet states, we need to solve for the equations of motion of these three-point correlators.
To do so, we must evaluate expressions such as
\begin{eqnarray}
\label{eq:phcomm}i\partial_t\ex{f_{a}^{\dag}(t)f_{b}(t) b_{-\vec{q}}} &=& \ex{[f_{a}^{\dag}(t)f_{b}(t) b_{-\vec{q}}, H_{\rm tot} - H(t)]} \nonumber\\
\label{eq:rescomm}i\partial_t\ex{f_{a}^{\dag}(t)d_{\ell}} &=& \ex{[f_{a}^{\dag}(t)d_{\ell} , H_{\rm tot} - H(t)]}.
\end{eqnarray}
Similar expressions are also needed for $i\partial_t \ex{f_{a}^{\dag}(t)f_{b}(t) b_{\vec{q}}^{\dag}}$ and $i\partial_t\ex{f_{b}(t)d_{\ell}^{\dag}}$.

The commutators in Eq.~(\ref{eq:rescomm}) generate many terms.
The corresponding calculation is straightforward, but somewhat tedious.
As above, the primary difference from the textbook case of a non-driven system \cite{Kira2012} is the appearance of sums over Floquet harmonic indices.

Mathematically, the crucial point is that the commutators in Eq.~(\ref{eq:rescomm}) give rise to higher order correlation functions such as $\ex{f^\dagger_af^\dagger_cf_bf_d\,b^\dagger_{\vec{q}}b_{\vec{q}'}}$ and $\ex{f^\dagger_af_bd^\dagger_\ell d_{\ell'}}$.
In the first case we split the averages into products of averages of fermionic and bosonic bilinear operators: $\ex{f^\dagger_af^\dagger_cf_bf_d\,b^\dagger_{\vec{q}}b_{\vec{q}'}} \approx \ex{f^\dagger_af_d}\ex{f^\dagger_cf_b}\ex{b^\dagger_{\vec{q}}b_{\vec{q}'}} - \ex{f^\dagger_af_b}\ex{f^\dagger_cf_d}\ex{b^\dagger_{\vec{q}}b_{\vec{q}'}}$, etc.
The fermionic averages involving system operators just give the polarizations $P^{a'}_a$ defined above.
We take the averages of the bosonic operators with respect to a thermal distribution with inverse temperature $\beta$: $\ex{b^\dagger_{\vec{q}}b_{\vec{q}'}} = \delta_{\vec{q}\vec{q}'}N(\omega_{\vec{q}})$, where $N(\varepsilon) = 1/(1 - e^{-\beta\varepsilon})$ and $\omega_{\vec{q}}$ is the frequency of bosonic mode $\vec{q}$.
Likewise, we split the averages involving reservoir degrees of freedom as $\ex{f^\dagger_af_bd^\dagger_\ell d_{\ell'}} \approx \ex{f^\dagger_af_b}\ex{d^\dagger_\ell d_{\ell'}}$.
For the Fermi reservoir, we take $\ex{d^\dagger_\ell d_{\ell'}} = \delta_{\ell\ell'} D(E_\ell)$, where $D(E)$ is the Fermi-Dirac function with temperature $T$ and chemical potential $\mu_{\rm res}$.
For brevity, below we use $D_\ell = D(E_\ell)$.

Through the above approximations we close the equation of motion hierarchy.
After splitting the averages on the right hand sides of Eq.~(\ref{eq:rescomm}), we integrate them from time 0 to $t$ to find the correlation functions $\ex{f_{a}^{\dag}(t)f_{b}(t) b_{\vec{q}}}(t)$ and $\ex{f_{a}^{\dag}(t)d_{\ell}}(t)$ needed as input for the equations of motion of the polarizations, Eqs.~(\ref{eq:ddtlevel1}), (\ref{eq:singletPhononTerm}) and (\ref{eq:singletReservoirTerm}).

To give an explicit example, we focus on one term which arises from the system-reservoir coupling,
\begin{widetext}
\begin{eqnarray}
\label{eq:fdCorrelatorEOM} i\partial_t\ex{f_{a}^{\dag}d_{\ell}}=E_{\ell}\ex{f_{a}^{\dag}d_{\ell}}+\sum_{b}\sum_{n}e^{-i\s E_{b}t}e^{-in\Omega t}\mathcal{J}_{\ell,b}^{(n)*}\left[P_{b}^{a}(1-D_{\ell})-(\delta_{ab}-P_{b}^{a})D_{\ell}\right].
\end{eqnarray}
 The calculation for other terms yields similar expressions. A straightforward formal integration, taking $\ex{f_{a}^{\dag}d_{b}}(t = 0) = 0$, then yields
\begin{eqnarray}
\ex{f_{a}^{\dag}d_{\ell}} & = & \frac{1}{i}\sum_{b}\sum_{n}\mathcal{J}_{\ell,b}^{(n)*}e^{-iE_{\ell}t}\int_{0}^{t}dze^{-i(\s E_{b}-E_{\ell})z}e^{-in\Omega z}\left[P_{b}^{a}(1-D_{\ell})-(\delta_{ab}-P_{b}^{a})D_{\ell}\right].
\end{eqnarray}
\end{widetext}
The next step is to introduce this result and its counterpart for $\ex{d^\dagger_{\ell}f_{a}}$ into Eq.~\eqref{eq:singletReservoirTerm}, for the contribution of the system-reservoir coupling to the evolution of the population $P^{a'}_a$, Eq.~(\ref{eq:ddtlevel1}).
Doing so, we obtain:
\begin{widetext}
\begin{eqnarray}
\nonumber \!\!\!\!\!\!\!\ex{[f_{a}^{\dag}f_{a'},H_{\rm tun}]} & = & \frac{1}{i}\sum_{\ell b}\sum_{mn}\mathcal{J}_{\ell,a'}^{(n)}\mathcal{J}_{\ell,b}^{(m)*} e^{i(\s E_{a'}-E_{\ell})t}e^{in\Omega t}\int_{0}^{t}dze^{-i(\s E_{b}-E_{\ell})z}e^{-im\Omega z}\left[P_{b}^{a}(1-D_\ell)-(\delta_{ab}-P_{b}^{a})D_{\ell}\right]\\
\label{eq:ResWithPol} & - & \frac{1}{i}\sum_{\ell b}\sum_{mn}\mathcal{J}_{\ell,a}^{(n)*}\mathcal{J}_{\ell,b}^{(m)}e^{-i(\s E_{a}-E_{\ell})t}e^{-in\Omega t}\int_{0}^{t}dze^{i(\s E_{b}-E_{\ell})z}e^{im\Omega z}\left[(\delta_{a'b}-P_{a'}^{b})D_{\ell}-P_{a'}^{b}(1-D_{\ell})\right].
\end{eqnarray}
\end{widetext}

Importantly, notice that the right hand side of Eq.~(\ref{eq:ResWithPol}) couples the evolution of the population $F_{a} = P^a_a$ to both the diagonal and off-diagonal polarizations $P_{b}^{a}$.
Thus in principle we do not have a closed set of equations for the populations alone.
In particular, for transient behavior (e.g.,~at early times when the driving is just switched on) such terms can not be ignored.

Close to the steady state, we may expect the off-diagonal polarizations (coherences) to be small under certain circumstances.
For a homogeneous system where the steady state maintains translational invariance, the polarizations in the steady state are diagonal in the electronic crystal momentum, $P^{k'\alpha'}_{k\alpha} \propto \delta_{kk'}$. Furthermore, coherences between the two Floquet bands can be suppressed in the steady state under suitable conditions, which are discussed at length in Appendix~\ref{appendix: rates}. These conditions are expected to be met for weak system-bath coupling, and we have verified that the steady states resulting from our simulations are indeed in this regime (see Appendix~\ref{appendix: rates}).

For strong system-bath coupling, the conditions discussed in Appendix~\ref{appendix: rates} might not be met, and a more complicated situation may arise. There, the particular form of system-bath coupling may try to drive the system towards specific states other than the Floquet states.
For example, relaxation may occur into the eigenstates of the non-driven system.
The competition between driving and relaxation may then lead to steady states featuring significant inter-Floquet-band coherences.

In this work we focus on the case of homogeneous steady states, with weak (but nonetheless realistic) system bath coupling.
We neglect all off-diagonal coherences, setting $P_{b}^{a} \propto \delta_{ab}$  in Eq.~(\ref{eq:ResWithPol}) and similarly for all other terms in the equations of motion.
Additionally, in the sums over Fourier harmonics we only keep the terms for which $n = m$; when the Floquet state populations evolve slowly on the timescale of the driving period, the terms with $n \neq m$ give rise to fast oscillations and thus produce negligible contributions.
With these two simplifications, the standard Markovian approximation yields the full Floquet kinetic equation:
\begin{eqnarray}
\partial_t F_{k \alpha}& = & I^{\rm ph}_{k\alpha} + I^{\rm rec}_{k\alpha} + I^{\rm tun}_{k\alpha},
\end{eqnarray}
with the collision integral for electron-phonon scattering given by
\begin{widetext}
\begin{eqnarray}
\!\!\!\!\!\!\!\!\!\!\!\!\!\!\!\!\!\!\!I^{\rm ph}_{k\alpha} &=& 
\label{eq:phononFinal}\frac{2\pi}{\hbar} \sum_{\alpha'\vec{q}}\sum_{n}|\mathcal{G}^{(n)}_{\alpha'\alpha}(k,q_x)|^{2}\left[F_{k-q_x\alpha'}\bar{F}_{k\alpha}N(\hbar\omega_{\vec{q}}) - F_{k\alpha}\bar{F}_{k-q_x\alpha'}(1+N(\hbar\omega_{\vec{q}}))\right]\delta(\s E_{k\alpha}-\s E_{k-q_x\alpha'} - \hbar\omega_{\vec{q}} - n\hbar\Omega)\ \ \ \ \ \\
& + & 
\nonumber\frac{2\pi}{\hbar}\sum_{\alpha'\vec{q}}\sum_{n}|\mathcal{G}^{(n)}_{\alpha'\alpha}(k,q_x)|^{2}\left[F_{k-q_x\alpha'}\bar{F}_{k\alpha}(1+N(\hbar\omega_{\vec{q}}))-F_{k\alpha}\bar{F}_{k-q_x\alpha'}N(\hbar\omega_{\vec{q}})\right]\delta(\s E_{k-q_x\alpha'} - \s E_{k\alpha} - \hbar\omega_{\vec{q}} + n\hbar\Omega).
\end{eqnarray}
Tunneling in and out of the Fermi reservoir is described by (see Eqs.~(\ref{eq:FFGR_Lead}) and (\ref{eq:I_Lead}) in the main text):
\begin{eqnarray}
I^{\rm tun}_{k\alpha} = \frac{2\pi}{\hbar} 
\sum_{\ell}\sum_{n}|\mathcal{J}_{\ell,k\alpha}^{(n)}|^{2}\left[\bar{F}_{k\alpha}D(E_\ell) - F_{k\alpha}(1-D(E_\ell))\right]\delta(\s E_{k\alpha} - E_\ell + n\hbar\Omega).
\label{eq:finalResultKineticEquation}
\end{eqnarray}
\end{widetext}
The collision integral corresponding to radiative recombination looks identical to that for electron-phonon scattering in Eq.~(\ref{eq:phononFinal}), with the matrix elements $\mathcal{G}_{\alpha'\alpha}^{(n)}(k,q_x)$ replaced by the appropriate ones for coupling to the electromagnetic environment.
In our model, the matrix element for coupling to bath photons is purely off-diagonal in the basis of the conduction and valence bands of the non-driven system. This model is motivated by the form of radiative transitions for electrons near $k=0$ in many experimentally relevant materials.   For simplicity,  we modeled recombination as ``vertical'' transitions~\cite{footnote:recombination}, giving  $G^{\rm rec}  = g^{\rm rec}(1-\delta_{\nu\nu'})\delta_{q_x,0}\delta_{k,k'}$.

According to our convention in Eq.~(\ref{eq:FloquetState}),  the fact that $G^{\rm rec}\propto (1-\delta_{\nu\nu'})$  requires a coupling between $\kett{\phi_{k\alpha}^{m}}$ and $\kett{\phi_{k\alpha'}^{m'}}$, where $m$ and $m'$ are separated by an odd integer for the case $\tilde{g}_\parallel = 0$.
Furthermore, the conservation of energy expressed by the delta function in Eq.~(\ref{eq:phononFinal}) requires $n$ to be {\it negative}.
Therefore, in our model recombination only acts through terms in Eq.~(\ref{eq:phononFinal}) with $n < 0$ odd.
The dominant contribution comes for $n = -1$ for weak driving.
Correspondingly, the emitted photon energy is large (on the order of the driving frequency), and hence we set all Bose occupation factors for photons to zero (i.e.,~only spontaneous emission is included).

Finally, to get the collision integral (\ref{eq:phononFinal}) into the form of Eq.~(\ref{eq:CollisionInt}) in the text, we integrate over the delta function to get the density of states for bosons with momentum $q_x$ parallel to the system.
This gives
\begin{widetext}
\begin{eqnarray}
I^{\rm ph}_{k\alpha} &=& 
\frac{2\pi}{\hbar} \sum_{k'\alpha'}\sum_{n}|\mathcal{G}^{(n)}_{\alpha'\alpha}(k,q_x)|^{2}\left[F_{k'\alpha'}\bar{F}_{k\alpha}N(\Delta\mathcal{E}_{n}) - F_{k\alpha}\bar{F}_{k'\alpha'}(1+N(\Delta\mathcal{E}_{n}))\right]\rho_{q_x}(\Delta\mathcal{E}_n)\\
\label{eq:phononFinalForReal} & + & 
\frac{2\pi}{\hbar}\sum_{k'\alpha'}\sum_{n}|\mathcal{G}^{(n)}_{\alpha'\alpha}(k,q_x)|^{2}\left[F_{k'\alpha'}\bar{F}_{k\alpha}(1+N(-\Delta\mathcal{E}_n))-F_{k\alpha}\bar{F}_{k'\alpha'}N(-\Delta\mathcal{E}_n)\right]
\rho_{q_x}(-\Delta\mathcal{E}_n),\nonumber
\end{eqnarray}
where in the above $q_x=k-k'$ and $\Delta\mathcal{E}_n = \s E_{k\alpha}-\s E_{k'\alpha'} - n\hbar\Omega$.
\end{widetext}

\section{System size scaling of transition rates}
\label{appendix: scaling}

In this section we discuss the scaling of the electron-boson scattering rates $W^{k'\alpha'}_{k\alpha}$ with the system size.
We will show that in the limit of a large system, the rates scale as $~\sim 1/L$. As we explain below, this  implies that both $\gamma^{\textrm{rec}}$ and $\Lambda^{\rm inter} = L\overline{W}^{\textrm{inter}}$, defined in Eq.~(\ref{eq:gamma_rec}) and the discussion below, are independent of system size. An important consequence is that the excitation density $n_{\textrm{steady}}$ is also \textit{independent} of the system size, as one would naturally expect. We first focus our discussion on the radiative recombination, i.e. interaction with a photon bath, and then explain how it can be easily applied also to a bath of phonons. To simplify the discussion, we illustrate the scaling using a non-driven toy model, but the discussion can be easily generalized for transition rate between Floquet states in a driven system.

We consider an electronic Bloch Hamiltonian of the form $H_0(k)=\left[2A\p{1-\cos(ka)}+E_{\textrm{gap}}\right]\sigma_z$. We define the Bloch states as $|k\alpha\rangle = \frac{1}{\sqrt{N}}\sum_{x=0}^{(N-1)a} e^{ikx}|x,\alpha\rangle$,
where $c,v$ correspond to the positive and negative eigenvalues of $\sigma_z$,  $a$ is the lattice constant and $L=Na$ is the electronic system size.

The electron photon interaction Hamiltonian, in the rotating wave approximation,  is given by $H_{\rm int}=\sum_{\vec{q}}H_{\rm int}(\vec{q})$, with
\begin{equation}
H_{\rm int}(\vec{q}) = \sum_{x} M_{\vec{q}}e^{i\vec{q}\cdot\vec{r}}\left( c^\dagger_{x, v}c_{x, c}b^\dagger_{\vec{q}} + c^\dagger_{x, c}c_{x, v} b_{-\vec{q}}\right) +h.c.,
\end{equation}
where in the above $c^\dagger_{x, \alpha}$ are creation and annihilation operators for Wannier states in the conduction and valence bands, and $|M_{\vec{q}}|$ depends on the volume of the electromagnetic environment as $|M_{\vec{q}}|\sim 1/\sqrt{V_{\textrm{env}}}$. Note that $H_{\rm int}(\vec{q})$ is diagonal in the lattice coordinate $x$. The rate for recombination from $|k, c\rangle$ to $|k', v\rangle$  is then given by
\begin{equation}
W^{k'v}_{kc} = \frac{2\pi}{\hbar} \sum_{\vec{q}}\left|M_{\vec{q}}\sum_{x}\frac{e^{i(k-k'+q)x}}{N}\right|^2\delta( E_{kc}- E_{k'v} - \hbar\omega_{\vec{q}}),
\label{eq:phononFinalSecB}
\end{equation}
where $\{E_{k\alpha}\}$ are the eigenenergies of $H_0(k)$.
Importantly, the photon momentum lives on a \textit{different reciprocal lattice} than the momenta of the electronic system, $q=\frac{2\pi}{L_{\textrm{env}}}n$. For simplicity, we drop the $\vec{q}$ dependence of $M_\vec{q}$. Summing over the \textit{transverse} photon momenta $\vec{q}_\perp$ yields a 2D density of states for the transverse modes with $q_x$ held fixed,
\begin{equation}
W^{k'v}_{kc}  = \frac{2\pi}{\hbar} \sum_{q_x}\frac{|M|^2}{N^2}\left|\sum_{x}e^{i(k-k'+q)x}\right|^{2}\rho_{q_x}( E_{kc}- E_{k'v}).
\end{equation}
 Note that $\rho_{q_x}( E_{kc}- E_{k'v})$ has dimensions of $\frac{1}{\textrm{Energy}}$ and scales as $L_{\textrm{env}}^2$. The photons emitted by the radiative recombination transition have a typical energy of $E_{\textrm{gap}}$, and therefore the corresponding photon momentum $\hbar q_* = E_{\textrm{gap}}/c$ plays an important role in the calculation of the rates. For simplicity, we set the density of states  $\rho_{q_x}( E_{kc}- E_{k'v})$ to be a constant $\rho_0^{(2D)}$ for $|q_x|\leq q_*$, and zero otherwise.
This gives
\begin{equation}
W^{k'v}_{kc}  = \frac{2\pi}{\hbar}\frac{M^2\rho_0^{(2D)}}{N^2} \sum_{q_x=-q_*}^{q_*}\left|\sum_{x}e^{i(k-k'+q)x}\right|^2.
\end{equation}
Assuming a large environment volume, we can write
\begin{equation}
W^{k'v}_{kc}  = \frac{2\pi}{\hbar}\frac{M^2\rho_0^{(2D)}L_{\textrm{env}}}{N^2} \int_{-q_*}^{q_*}\frac{dq}{2\pi}\left|\frac{1-e^{iq L}}{1-e^{i(k-k'+q)a}}\right|^2\!\!.
\label{eq:WFinal}
\end{equation}
Recalling that $M\sim 1/\sqrt{V_\textrm{env}}$ we see that the factor $M^2\rho_0^{(2D)}L_{\textrm{env}}$ is \textit{independent of environment size}.

The calculation now amounts to evaluating the integral in Eq.~(\ref{eq:WFinal}).
Using integers to represent momenta as in $k=\frac{2\pi}{L}n$, we denote this integral by $g_N(n-n')$, where the $N$ subscript denotes the fact that the integral depends on the system size $L=Na$. We are interested in the scaling of this integral with $N$. We define the dimensionless variable $\tilde{q}=q L$, and divide by $N^2$ for later convenience, whereby the integral becomes
\begin{equation}
\frac{g_N(m)}{N^2}=\frac{1}{L}\int_{-q_*Na}^{q_*Na}d\tilde{q}\frac{\sin^2(\tilde{q}/2)}{N^2\sin^2(\frac{1}{2N}\left[2\pi m+\tilde{q}\right])}.
\label{eq: g final}
\end{equation}
Note that in the prefactor on the right hand side above, $L$ is the \textit{electronic system} size. In order for $W^{k'v}_{kc}\sim 1/L$, which guarantees that e.g. $n_{\textrm{steady}}$ remains independent of system size, one has to have that $\frac{g_N(m)}{N^2}\sim 1/L$.

In the following, we assume $q_* a=\frac{E_{gap}a}{\hbar c} \ll 1$, which means the photon wavelength is much larger then the lattice spacing of the system.
We furthermore consider the limit where the system size is larger than the photon wavelength, $N \gg 1/(q_* a)$. We start again from Eq.~(\ref{eq: g final}).
Clearly, for $g_N(m)/N^2$ to be of order $1/L$, we must have $2\pi|m| \lesssim  N q_* a$, which guarantees that the integral picks the contribution where the $\sin$ function in the denominator of (\ref{eq: g final}) vanishes. Physically, this corresponds to the requirement that $|k-k'|\lesssim q_*$.

We now need to check how the integral in Eq.~(\ref{eq: g final}) scales with $N$. We do this explicitly for $m=0$; the result can be generalized for any $2\pi|m| \lesssim  N q_* a$.
We break the integral into three  integration regions: (1)  $\left[-\sqrt{q_*Na},\sqrt{q_*Na}\right]$  (2) $\left[\sqrt{q_*Na},q_*Na\right]$ and (3) $\left[-q_*Na,-\sqrt{q_*Na}\right]$. In region (2), we can give an upper bound to the integral by
\begin{equation}
 \int_{\sqrt{q_*Na}}^{q_*Na}d\tilde{q}\frac{1}{N^2\sin^2(\sqrt{q_*a/4N})} \xrightarrow{N\rightarrow\infty} C,
\label{eq: integral bound}
\end{equation}
where $C$ is a constant. The same result applies to the integral in region (3). In region (1), we expand the denominator to obtain
\begin{equation}
\int_{-\sqrt{q_*Na}}^{\sqrt{q_*Na}}d\tilde{q}\frac{\sin^2(\tilde{q}/2)}{\frac{1}{4}\tilde{q}^2}\left(1-\frac{\tilde{q}^2}{12N^2}+...\right).
\label{eq: g final3}
\end{equation}
The first term in the above expansion clearly gives an order $1$ contribution, while the rest of the terms vanish in the limit of large $N$.

The result of the above analysis is that $g_N(m)\approx\frac{N^2}{L}\bar{g}(m)$, where $\bar{g}(m)$ is independent of system size. The full expression for $g_N(m)$ may contain  terms that scale slower than $N^2/L$ with the system size.
Finally, inserting this result this back into Eq.~(\ref{eq:WFinal}), we arrive at the scaling  $W^{k'c}_{kv}\sim 1/L$ of rates with the system size, in the limit of a large system.

Putting this into the definition of the \textit{total rate} of recombination out of the state $\kett{k,c}$, defined by $\mathcal{W}_k^{rec}= \sum_{k'}W^{k'v}_{kc}$, we get
 \begin{eqnarray}
\mathcal{W}_k^{rec}=L \int \frac{dk'}{2\pi} W^{k'v}_{kc} \approx L \frac{E_{gap}}{\pi\hbar c} W^{kv}_{kc}.
\label{eq: Wrecappendix}
\end{eqnarray}
 Therefore $\mathcal{W}_k^{rec}$ is \textit{independent} of system size, as promised. Likewise, the \textit{total rate density}, $\gamma^{\textrm{rec}}=\frac{1}{L} \sum_{k,k'}W^{k'v}_{kc} =\int\frac{dk}{2\pi} \mathcal{W}_k^{rec}$ is \textit{independent} of system size. Note that the factor  $\frac{E_{gap}}{\pi\hbar c}$ in Eq.~(\ref{eq: Wrecappendix}) accounts for the photon density of states in the longitudinal direction, whereby $\rho^{\textrm{3D}}=\frac{E_{gap}}{\pi\hbar c}\rho^{\textrm{2D}}$.

\subsection{Scaling of phonon matrix elements}

The treatment of the phonon matrix elements follows along the same lines as above. 
Let us treat the longitudinal size (along the direction of the one dimensional system) of the phonon bath to be equal to the system size $L$. 
This ensures the conservation of crystal momentum, $k' + q = k$.
 We start from the analogue of Eq.~(\ref{eq:phononFinalSecB}), for phonon scattering rates. Performing the sum over $x$ we obtain a factor of $N^2\delta_{k'-k,q_x}$. In this case, however, the factor $M^2\rho_0^{2D}$ scales as $\sim 1/L$. Therefore, $W^{k'\alpha'}_{k'\alpha}\sim1/L$.

\section{Numerical simulations}
\label{appendix: parameters}

In our numerical simulations, the steady state distributions were obtained by direct evolution of Eq.~\eqref{eq:KineticEq}. The results
are independent of the initial distribution $\{F_{k\alpha}(t=0)\}$.

In Sec.~\ref{sec:bosons} we discussed the square root dependence of the excitation density $n_{\textrm{steady}}$ on the ratio $\pi\overline{\mathcal{W}}^{\textrm{rec}}/k_R W^{\textrm{inter}}$. This behaviour was observed in our numerical simulations, as shown in Fig.~\ref{fig:Results1}. Below we discuss how each factor in the above ratio was calculated from the numerical data. For recombination, the quantity $\overline{\mathcal{W}}^{\textrm{rec}}$ is calculated by averaging the quantity $\mathcal{W}^{\textrm{rec}}_k$ [defined below Eq.~(\ref{eq:gamma_rec})],  in the interval $\left[-k_R,k_R\right]$. Since $\mathcal{W}^{\textrm{rec}}_k$ becomes negligible far outside of this interval, we define $\overline{\mathcal{W}}^{\textrm{rec}}=\pi/(k_R L)\sum_k \mathcal{W}^{\textrm{rec}}_k$, with

\begin{equation}
{\mathcal{W}_{k}^{\textrm{rec}} = \frac{2\pi}{\hbar}|g^{\textrm{rec}}|^2\rho_0 \sum_{n=1}^\infty\Big\vert\sum_{m}\qb{\phi_{k+}^{m-n}}{kv}\qb{kc}{\phi_{k-}^{m}}\Big\vert^2}.
\label{eq: wrec}
\end{equation}

To estimate the typical \textit{interband} scattering rate from phonons, we evaluate $\overline{W}^{\rm inter}$ by taking an average interband phonon scattering rate near the resonances $\pm k_{R}$. The explicit form for $\overline{W}^{\rm inter}$ we used is
\begin{equation}
\overline{W}^{\textrm{inter}}  =  \frac{2}{N_{\epsilon}^{2}}\sum_{k= k_{R}-\epsilon}^{k_{R}+\epsilon}\;\sum_{k'= k_{R}-\epsilon}^{k_{R}+\epsilon}W_{k+}^{k'-}
\label{eq: winter}
\end{equation}

In the above, the rates $W_{k+}^{k'-}=W_{k+}^{k'-}(0)$, defined in Eq.~(\ref{eq:FFGR}), corresponds to interband transitions from the upper to the lower Floquet band, through the energetically allowed phonon \textit{emission} processes in the model that we studied numerically.  The factor of $2$ comes from summing over transitions at momenta around  $\pm k_R$.   Furthermore, in Eq.~(\ref{eq: winter}), we denote by $N_{\epsilon}$  the number of $k$-points corresponding
to the region near $k_{R}$ set by $\epsilon=\frac{\pi}{10a}$. The
excitation density, normalized to the thermal density, is fitted  using nonlinear least squares, to
the form $\log_{10}(\frac{n_{e}}{n_{th}})=p\log_{10}\left(\frac{k_R}{\pi}\frac{\mathcal{W}^{\textrm{rec}}}{W^{\textrm{inter}}}\right)+(b-\log_{10}(n_{th}))$
to obtain $p=0.49$ and $b=0.95$ with a standard error of $0.001$
in the region away from saturation.

\section{Scattering rates in the steady state}
\label{appendix: rates}

Throughout this paper we have used the single particle Floquet states to describe the steady state of the system.
We focused on a regime in which the steady state approximately yields a single particle density matrix which is diagonal in the basis of Floquet states, (i.e., the ``off diagonal'' correlations of the form $\langle f_{k\alpha}(t)f_{k'\beta}\rangle$, and higher order correlations, are negligible for $k \neq k'$ and/or $\alpha \neq \beta$).
In this regime, the steady state of the system can be efficiently described in terms of the occupation of single particle Floquet states.

Clearly, in order for the system to be in the ``diagonal'' regime described above, the lifetimes of the single particle Floquet states need to be much longer than the driving period.
However, for Floquet-Bloch states near the resonance momenta $k_R$, we expect a more stringent criterion to be necessary.
To see why, recall that near the resonance momenta, the Floquet states are approximately equal amplitude coherent superpositions of the original (non-driven) conduction and valence bands.
When initialized to one of these original eigenstates, the system oscillates between the conduction and valence band with a Rabi frequency equal to the Floquet gap $\Delta_{k_R}$, which typically can be expected to be smaller than the driving frequency.
For these oscillations (and hence the Floquet states) to be well resolved, and hence for our approach to be valid,
the scattering rates between Floquet states in the steady state must be smaller than the Floquet gap.

Our model includes three scattering mechanisms: electron-phonon interaction, radiative recombination, and the coupling to the Fermi reservoir.
Typically in semiconductors, radiative recombination rates are on the order of $1\, \textrm{ns}^{-1}$.
The Rabi frequency depends on driving power, and may be on the order of $ 0.5\, \textrm{ps}^{-1}$ or even larger
in experimentally accessible setups (see, e.g.,~Refs.~\onlinecite{Wang2013,Vu}). Therefore, the contribution from recombination to the scattering rate from a state can be significantly smaller than the Floquet gap. Phonon scattering rates can be appreciably larger. An order of magnitude for the ``bare'' scattering rate, (not suppressed by Pauli blocking), is on the order of $1\, \textrm{ps}^{-1}$.  However, as we discuss below, Pauli blocking and phase space considerations can significantly reduce the scattering rates for populated Floquet states. Finally, we study how the steady state evolves when we vary a coupling strength to a Fermi reservoir. Importantly, even when the coupling to the reservoir is strong enough to significantly suppress the densities of excited electrons and holes, it may induce tunneling rates to and from the reservoir which are still significantly smaller than the ``bare'' rate for scattering from phonons.

When the driven system's density is close to half filling, our results indicate that the steady state of the system resembles that of a Floquet insulator with an added density of excited electrons and holes. Filled states in the lower Floquet band with $|k|\gg k_R$ have suppressed scattering rates from \textit{phonons} due to Pauli blocking. Furthermore, for small $\tilde{g}_\parallel$ or for weak driving, the radiative recombination rate out of these states is strongly suppressed.  Likewise, since these states are coupled to filled states of the filtered reservoir, the reservoir does not introduce any further scattering out of these states. Therefore, overall we expect negligibly small scattering rate in the lower Floquet band for $|k|\gg k_R$.

There are two types of momentum regions in the Brillouin zone with a non-negligible occupation of electrons that require a more careful treatment of the scattering rates. Regions of the first type are those near the resonance momenta, $\pm k_R$, where the distributions of excited electrons $n_e(k)=\langle f^\dagger_{k+}f_{k+} \rangle$ and holes $n_h(k)=1-\langle f^\dagger_{k-}f_{k-} \rangle$ are localized.  In the regime of low densities of electrons and holes studied in this paper, the distributions $n_e(k)$ and $n_h(k)$ are far from a \textit{degenerate} Fermi gas due to the temperature of the phonon and Fermi reservoirs, $k_BT=0.1\Delta_{k_R}$. Therefore, scattering of electrons and holes is not significantly affected by Pauli blocking in these regions. However, since $n_e(k)$ and $n_h(k)$ are highly localized around the minima and maxima of the Floquet bands, \textit{intraband} scattering from acoustic phonons is suppressed due to reduced phase space for these processes. The \textit{interband} phonon scattering rates are also reduced because of phase space arguments: recall that an electron in the upper Floquet band can only relax to momentum states around $k_R$ in the lower Floquet band; these are the considerations which led us to the square root behavior in Eq.~(\ref{eq:rho_steady}).

The second region comprises momenta between the two resonant momenta, $|k|\leq k_R$. This is the region in the Brillouin zone where the Floquet bands have an inverted character relative to the original valence and conduction bands. This, therefore, is the momentum range where most of the radiative recombination occurs. Since recombination transfers electrons from full states in the lower Floquet band to empty states in the upper Floquet band, the total scattering rate of electrons in this region is expected to be set by the recombination rate.

In order to compare the scattering rates in the steady state to the Floquet gap, we numerically evaluate the \textrm{total} scattering rate in the steady state of the system. Denoting by $\Gamma^{\textrm{tot}}_{k\alpha}$ the inverse lifetime of a test particle which we initialize in a Floquet state $\alpha$ at momentum $k$, we get
\begin{widetext}
\begin{equation}
\Gamma^{\textrm{tot}}_{ k \alpha}= \sum_{k'\alpha'}\left\{W_{k\alpha}^{k'\alpha'}\left[1+N(\mathcal{E}_{k\alpha}-\mathcal{E}_{k'\alpha'})\right]+W_{k'\alpha'}^{k\alpha}N(\mathcal{E}_{k'\alpha'}-\mathcal{E}_{k\alpha})+\mathcal{W}_{k}^{\textrm{rec}}\delta_{kk'}\delta_{\alpha -}\delta_{\alpha' +}\right\}\left(\bar{F}_{k'\alpha'}\right)+\Gamma^0_{k\alpha}\left[1-D(\mathcal{E}_{k\alpha})\right].
\label{eq: Gamma tot}
\end{equation}
\end{widetext}

In the above, the first two terms correspond to electron-photon scattering. In the first term, the rates $W_{k\alpha}^{k'\alpha'}=W_{k\alpha}^{k'\alpha'}(0)$, defined in Eq.~(\ref{eq:FFGR}), correspond to the energetically allowed phonon \textit{emission} processes in the model we studied numerically. Note that $W_{k\alpha}^{k'\alpha'}=0$ when $\mathcal{E}_{k}-\mathcal{E}_{k'}<0$, due to the requirement for a nonzero density of states for the phonons. The second term in Eq.~(\ref{eq: Gamma tot}) corresponds to phonon \textit{absorption}, described by the rates $W^{k\alpha}_{k'\alpha'}=W^{k\alpha}_{k'\alpha'}(0)$ which vanish when $\mathcal{E}_{k}-\mathcal{E}_{k'}>0$. Furthermore, in Eq.~(\ref{eq: Gamma tot}),
$\mathcal{W}_{k}^{\textrm{rec}}$ [see Eq.~(\ref{eq: wrec})] is the radiative recombination rate out of the state $k,-$ to the state $k,+$  (recall that we model recombination with ``vertical'' transitions). The results are given in Fig.~\ref{fig: rates}, which shows $\Gamma^{\textrm{tot}}_{k\alpha}$ normalized to the Floquet gap $\Delta_{k_R}$ in both Floquet bands, for two representative values of the coupling strength to an energy filtered Fermi reservoir. The chemical potential of the reservoir is placed in the middle of the Floquet gap. In the same figures, we plot the numerically obtained distributions of electrons $n_e(k)$ and holes $n_h(k)$.

\begin{figure}[t]
\begin{center}
\includegraphics[width=1.0\columnwidth]{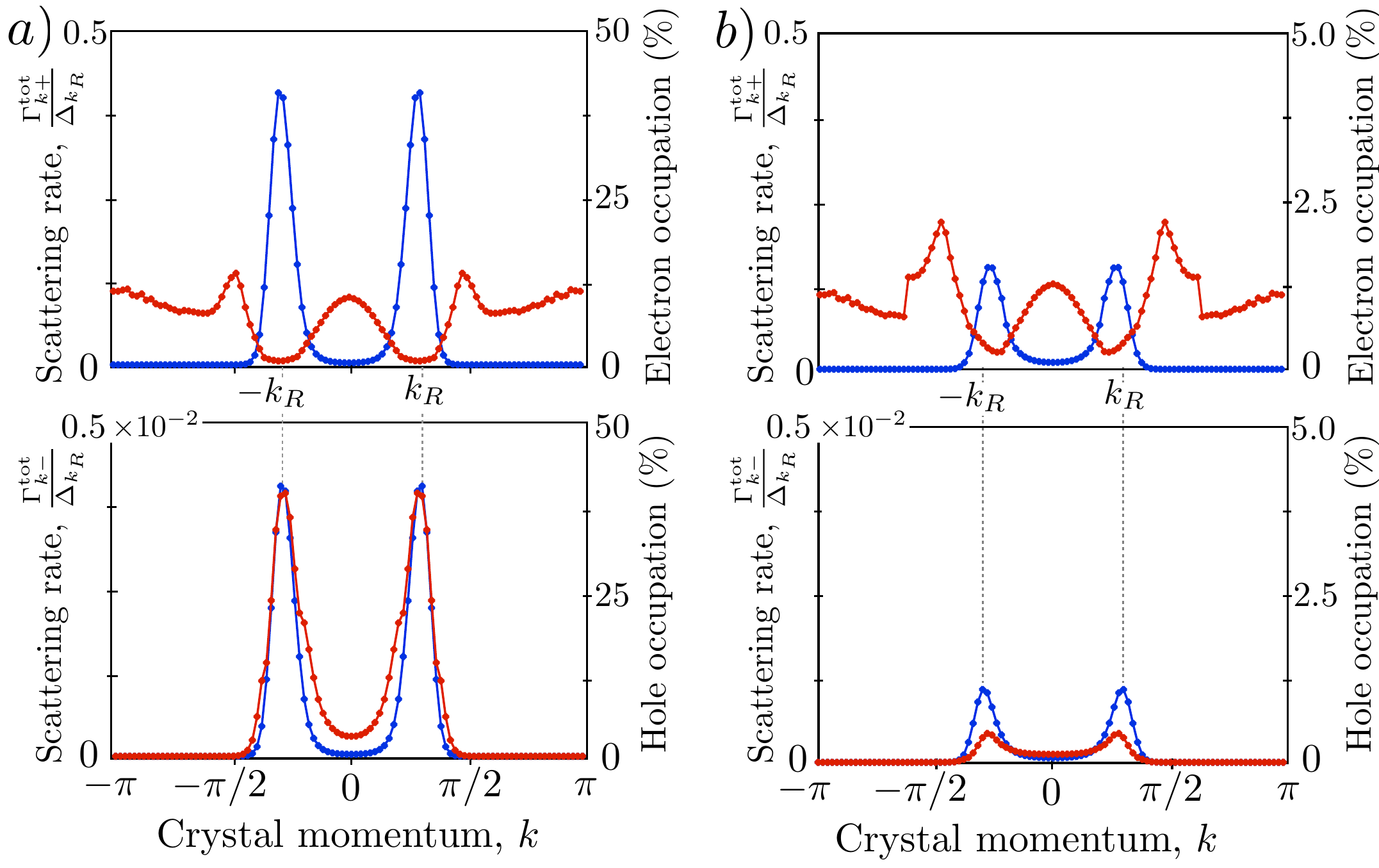}
\caption{Scattering rates (red), $\Gamma_{k\alpha}^{\textrm{tot}}$, [see Eq.~(\ref{eq: Gamma tot})] in the steady state of the system. The top (bottom) plot corresponds to the upper (lower) Floquet band. Also shown (blue) are the distributions $n_e(k)$ an $n_h(k)$ in each Floquet band. In panel (a), we show the case of half-filling, with no reservoir coupling (corresponding to Fig.~\ref{fig:Results1} in the main text) while in (b) we take $\log_{10}(\Upsilon)=3$. Other model parameters are the same as for the green (middle) curve in Fig.~\ref{fig:Results1}. Note the enhanced scale for the rates in the bottom plots, and the enhanced scale for the distributions in panel (b).}
\label{fig: rates}
\end{center}
\end{figure}

Let us first examine the situation when the system is not coupled to the Fermi reservoir, Fig.~\ref{fig: rates}a. Consider the scattering rates in the lower Floquet band. For values of $|k|$ which are significantly larger than $k_R$, the scattering rate vanishes; as described above this is due to Pauli blocking, which prohibits scattering from phonons. Other scattering mechanisms are absent in this momentum region, as explained above. In the momentum region $|k|<k_R$, recombination is active.  For the simulations shown, the ``bare'' recombination rate (defined without taking into account the occupations $F_{k\alpha}$)  is taken to be $\mathcal{W}_k^{\textrm{rec}}=3\times 10^{-5} \Delta_{k_R}$, in line with experimentally accessible parameter regimes.  Therefore, a small nonzero $\Gamma^{\textrm{tot}}_{k\alpha}$, set by the recombination rate, can be seen in Fig.~\ref{fig: rates}a. Finally, for momenta $|k|\approx k_R$, more significant scattering rates can be observed, due to the nonzero density of holes and the possibility for scattering from phonons. However, due to phase space restrictions, the scattering rate is suppressed relative to its maximal possible value (see below), and therefore it is significantly smaller than the Floquet gap.

Next, we examine the rates in the upper Floquet band. In this band, momentum states with $|k|\gg k_R$ and $|k|\ll k_R$ are mostly unoccupied, and therefore a test particle initialized in these momentum states is expected to have more significant scattering from phonons. The maximal scattering rate in Fig.~\ref{fig: rates} is predominantly due to phonon scattering. Importantly, compared with this maximal scattering rate, the scattering rates at momenta $k\approx k_R$ are significantly suppressed due to reduced phase space for phonon scattering, as argued above.

Finally, we consider the scattering rates when the system is connected to an energy filtered Fermi reservoir. We note that increasing the coupling strength to the Fermi reservoir increases the scattering rates in the upper Floquet band for states with $|k| \lesssim k_R$. This is a result of the significant original conduction band component in these Floquet states, which is coupled to predominantly empty reservoir states (up to thermally induced corrections). The rates in the lower Floquet band are only weakly affected, as this band is coupled to predominantly filled reservoir states.

\section{Particle-hole asymmetry due to an energy filtered lead}
\label{appendix: asymmetry}

In Sec.~\ref{subsec:steadyStateFiltered}, we studied the situation where the system is coupled to an energy filtered Fermi reservoir with its chemical potential $\mu_{\textrm{res}}$ placed in the middle of the Floquet gap.
Here we generically find a nonzero difference between the densities of excited electrons and holes, $\Delta n = n_e-n_h$, despite the symmetry of the Floquet band structure.
This highlights one of the interesting features of driven systems, where steady state level occupations depend {\it both} on the state of the bath, as well as on the detailed form of the system-bath coupling.

To see how a nonzero $\Delta n$ arises, we first note that in order to have $\mu_{\textrm{res}}$ in the middle of the Floquet gap, we aligned it with the resonant energy $\frac12 \hbar\Omega$ in the conduction band. This placement manifestly breaks the particle hole symmetry of the system. Consider how the Floquet states $|\psi_{k\alpha}(t)\rangle$ are coupled to the reservoir, as depicted in Fig.~\ref{fig:filter}. Because the reservoir is energy filtered, within our convention for defining the quasi-energy zone 
the available reservoir states only couple to the $|\phi_\pm^0\rangle$ harmonics of the Floquet states, i.e., only the rates with $n=0$ in Eq.~(\ref{eq:FFGR_Lead}) are nonzero.  Consider the situation now for weak driving or small $\tilde{g}_\parallel$. The $n=0$ harmonics are then predominantly formed from conduction band components, $|\phi^0_{k\pm}\rangle \approx | c ,k \rangle$. Therefore, in the lower Floquet band, only states with momenta $ -k_R \lesssim k \lesssim k_R$ have an appreciable $|\phi^0_{k-}\rangle$ component, and therefore only those are coupled to the reservoir. The situation in the upper Floquet band is reversed: only states with $|k| \gtrsim k_R$ have an appreciable $|\phi^0_{k+}\rangle$ component, and are therefore coupled to the reservoir.

\begin{figure}[t]
\begin{center}
\includegraphics[width=1.0\columnwidth]{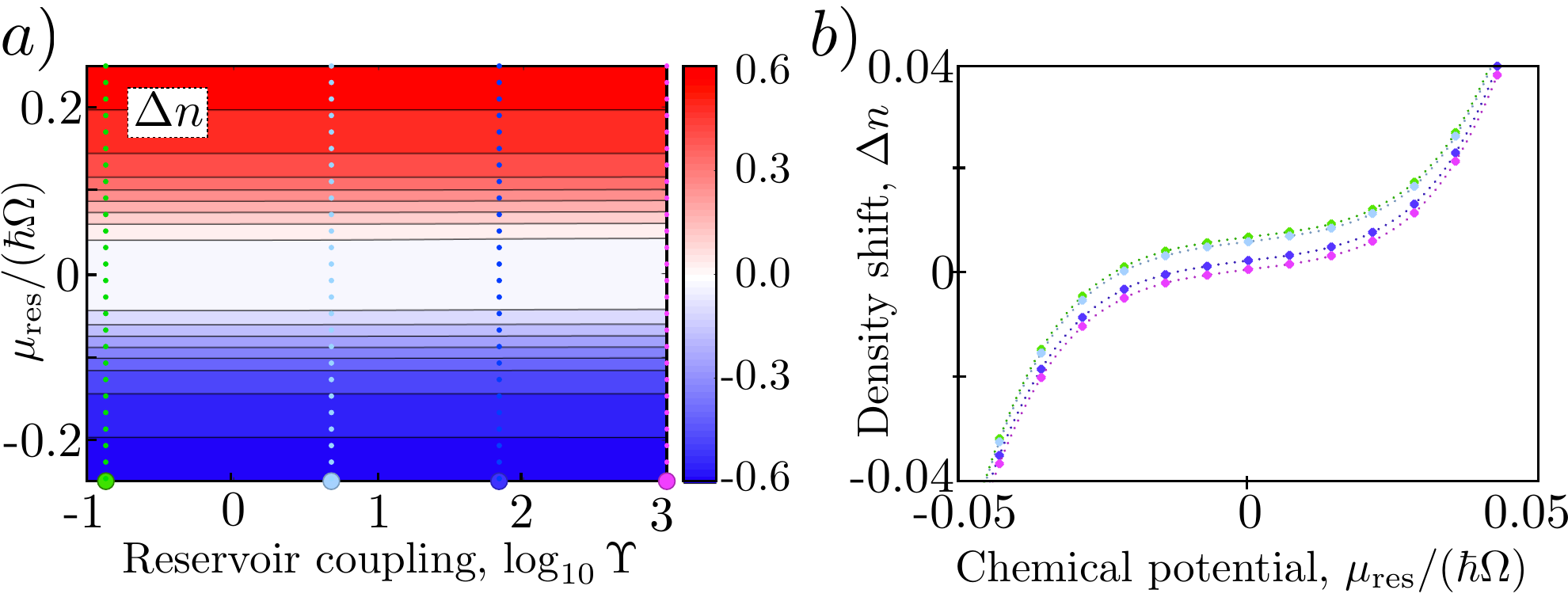}
\caption{The offset density $\Delta n$. Panel (a) shows $\Delta n$ as a function of the reservoir chemical potential $\mu_{\textrm{res}}$, and the coupling of the Fermi reservoir $\log_{10}\Upsilon$. Incompressible behavior can be seen when $\mu_{\textrm{res}}\approx 0$. (b) Vertical cuts of panel (a), showing $\Delta n$ as a function of $\mu_{\textrm{res}}$ for several coupling strengths to the Fermi reservoir, with values indicated by the dashed lines in panel (a). The small slope can be attributed to activated behavior due to the finite temperature of the reservoir.}
\label{fig: Delta n}
\end{center}
\vspace{-0.1 in}
\end{figure}

From the above considerations, we see that the rates $\Gamma^0_{k+}$ and $\Gamma^0_{k-}$ are not
equal in the regions around the resonance momenta $k_R$. As a consequence, the following rates are not equal: (1) the rate for excited electrons in the distribution $n_e(k)$ to tunnel to the reservoir, and (2) the rate for electrons in the reservoir to tunnel into empty states of the distribution $n_h(k)$. The difference between these rates leads to the non-zero value of $\Delta n$.

To conclude this section, we examine how $\Delta n$ behaves as $\mu_{\textrm{res}}$ is shifted away from the middle of the Floquet gap. In Sec.~\ref{subsec:steadyStateFiltered} we discussed
an incompressible behavior of the system.
Specifically, Fig.~\ref{fig:Results3}c showed that $\bar{n} = n_e + n_h$, which characterizes the number of ``free carriers,''  is unchanged by small shifts of $\mu_{\textrm{res}}$ around the middle of the Floquet gap.  
Here, we complement this result by plotting the
behavior for $\Delta n$ in Fig.~\ref{fig: Delta n}. The small slope of $\Delta n( \mu_{\textrm{res}})$ can be attributed to an activated behavior due to the finite temperature of the reservoir. Note that $|\Delta n|$ decreases as the coupling to the fermionic reservoir is increased. This is expected as $|\Delta n| < |\bar{n}|$.

\bibliographystyle{apsrev}
\bibliography{manuscript150209}

\begin{thebibliography}{63}
\bibitem[{foo({\natexlab{a}})}]{footnote:excitations}
\bibinfo{note}{While there is no notion of a ground state for the driven
  system, we use the term ``excitation'' over the ideal steady state to make
  connection to the analogous states of equilibrium systems.}
\bibitem[{foo({\natexlab{c}})}]{footnote:twoPhotonCondition}
\bibinfo{note}{The condition $E_{k} < 2\hbar\Omega$ ensures that the two photon
  resonance condition is never satisfied for a two-band system.}
\bibitem[{foo({\natexlab{d}})}]{footnote:recombination}
\bibinfo{note}{For systems larger than the photon wavelength, non ``vertical''
  recombination rates $W_{k-}^{k'+}$ may be non zero. Here we use a single rate
  $W_{k-}^{k+}$ to account for the combined contribution of all processes, as
  $\sum_{k '}W_{k-}^{k'+}$ is expected to be independent of system size (see
  Appendix~\ref{appendix: scaling}).}
\bibitem[{foo({\natexlab{e}})}]{footnote:smallPhoton}
\bibinfo{note}{We assume that the photon density of states at energies on the
  order of the Floquet gap $\Delta_{k_R}$ is negligible, and only consider
  emission of photons with energies $\hbar\omega \gtrsim E_{\rm gap}$.}
\bibitem[{foo({\natexlab{f}})}]{footnote:sqrt}
\bibinfo{note}{Note that $\overline{\mathcal{W}}^{\rm rec}$ is size-independent
  (see Appendix~\ref{appendix: scaling}), and therefore the density $n_{\rm
  steady}$ is also size-independent.}
\bibitem[{foo({\natexlab{g}})}]{footnote:widebandwidth}
\bibinfo{note}{In practice, the density of state is taken to be nonzero only in
  a window of width $11\Omega$ placed symmetrically around $E=0$.}
\bibitem[{foo({\natexlab{h}})}]{footnote:filterDriving}
\bibinfo{note}{If this criterion is not met, photon assisted tunneling rates
  will not be strictly suppressed, but rather will be controlled by the
  amplitude at which the filter is driven.}
\bibitem[{foo({\natexlab{j}})}]{footnote:finiteTemp}
\bibinfo{note}{When the reservoir is at finite temperature activated behavior
  can be expected as thermal excitations in the lead overlap with Floquet bands
  of the system.}
\end{thebibliography}

\end{document}